\documentclass[twocolumn,preprintnumbers,prb,superscriptaddress,amsmath,amssymb,amsfonts]{revtex4-2}
\usepackage{graphicx}  % needed for figures
\usepackage{color}
\usepackage[colorlinks,bookmarks=true,citecolor=blue,linkcolor=blue,urlcolor=blue, breaklinks=true]{hyperref}
\usepackage{times}
\usepackage{float}
\usepackage{gensymb}
\newcommand{\bolK}{\mathbf{K}}
\newcommand{\bolk}{\mathbf{k}}
\newcommand{\bolB}{\mathbf{B}}

\newcommand{\bolD}{\mathbf{D}}
\newcommand{\bolS}{\mathbf{S}}
\newcommand{\bolT}{\mathbf{T}}

\newcommand{\be}{\text{e}}
\newcommand{\SCBO}{\text{SrCu$_2$(BO$_3$)$_2$}}

\graphicspath{{./figures/}}
\begin{document}
%%%%%%%%%%%%%%%%%%%%%%%%%%%%%%%%%%%%%%%%%%%%%%%%%%%%%%%%%%%%
%%%%%%%%%%%%%%%%%% Title of paper %%%%%%%%%%%%%%%%%%%%%%%%%%%%%%%%%%%
\title{Intrinsic suppression of topological thermal Hall effect in an exactly solvable quantum magnet}

\author{S. Suetsugu}
\affiliation{Department of Physics, Kyoto University, Kyoto 606-8502, Japan}
\author{T. Yokoi}
\affiliation{Department of Physics, Kyoto University, Kyoto 606-8502, Japan}
\author{K. Totsuka}
\affiliation{Yukawa Institute for Theoretical Physics, Kyoto University, Kyoto 606-8502, Japan}
\author{T. Ono}
\affiliation{Department of Physics, Kyoto University, Kyoto 606-8502, Japan}
\author{I. Tanaka}
\affiliation{Department of Physics, Kyoto University, Kyoto 606-8502, Japan}
\author{S. Kasahara}
\affiliation{Department of Physics, Kyoto University, Kyoto 606-8502, Japan}
\affiliation{Research Institute for Interdisciplinary Science, Okayama University, Okayama 700-8530, Japan}
\author{Y. Kasahara}
\affiliation{Department of Physics, Kyoto University, Kyoto 606-8502, Japan}
\author{Z. Chengchao}
\affiliation{Department of Energy and Hydrocarbon Chemistry, Kyoto University, Nishikyo-ku, Kyoto, 615-8510, Japan}
\author{H. Kageyama}
\affiliation{Department of Energy and Hydrocarbon Chemistry, Kyoto University, Nishikyo-ku, Kyoto, 615-8510, Japan}
\author{Y. Matsuda}
\affiliation{Department of Physics, Kyoto University, Kyoto 606-8502, Japan}

\date{\today}

\begin{abstract}
In contrast to electron (fermion) systems, topological phases of charge neutral bosons have been poorly understood
despite recent extensive research on insulating magnets.
The most important unresolved issue is how the inevitable inter-bosonic interactions influence the topological properties.
It has been proposed that the quantum magnet SrCu$_2$(BO$_3$)$_2$ with an exact ground state serves as an ideal platform for this investigation,
as the system is expected to be a magnetic analogue of a Chern insulator with electrons replaced by bosonic magnetic excitations (triplons).
Here, in order to examine topologically protected triplon chiral edge modes in SrCu$_2$(BO$_3$)$_2$, we measured and calculated the thermal Hall conductivity $\kappa_{xy}$.
Our calculations show that the sign of $\kappa_{xy}$ is negative, which is opposite to the previous calculations, and its magnitude is 2$\pi$ times smaller.
No discernible $\kappa_{xy}$ was observed, which is at most 20-30\% of our calculations if present.
This implies that even relatively weak inter-particle interactions seriously influence the topological transport properties at finite temperatures. 
These demonstrate that, in contrast to fermionic cases, the picture of non-interacting topological quasi-particles 
cannot be naively applied to bosonic systems, 
calling special attention to the interpretation of the topological bosonic excitations reported for various insulating magnets.
\end{abstract}

%\keywords{}

%%%%%%%%%%%%%%%%%%%%%%%%%%%%%%%%%%%%%%%%%%%%%%%%%%%%%%%%%%%%
\maketitle

%%%%%%%%%%%%%%%%%  Main text %%%%%%%%%%%%%%%%%%%%%%%%%%%%%%%%%%%%%%
%%%%%%%%%%%%%%%%%%%%%%%%%%%%%%%%%%%%%%%%%%%%%%%%%%%%%%%%%%%%
\section{Introduction}
%%%%%%%%%%%%%%%%%%%%%%%%%%%%%%%%%%%%%%%%%%%%%%%%%%%%%%%%%%%%
The discoveries of electronic topological materials having gapped bulk excitations and topologically protected gapless edge states have led to a quest for similar effects in systems with quasiparticles obeying different quantum statistics.
In the past few years, quantum insulating magnets with Dzyaloshinskii-Moriya (DM) interaction have attracted renewed interest 
because they have raised the prospect of harboring topologically protected edge states of bosonic spin excitations \cite{PhysRevB.87.144101,PhysRevB.89.134409,PhysRevLett.115.147201}
(for recent developments, see, e.g., Ref. \cite{McClarty-review-21} and the references cited therein).
As these excitations are charge neutral, they do not respond to an electromagnetic field but can carry heat 
and potentially exhibit the thermal Hall effect \cite{PhysRevLett.104.066403,PhysRevLett.106.197202} 
without resorting to the Lorentz force.
Recently, finite thermal Hall conductivity $\kappa_{xy}$ has been experimentally resolved in several insulating magnets 
on geometrically frustrated lattices \cite{onose2010observation,PhysRevB.85.134411,hirschberger2015large,PhysRevLett.115.106603,watanabe2016emergence,PhysRevLett.121.097203,akazawa2020thermal},
which has been interpreted as the predicted topological thermal Hall effect of bosonic spin excitations.
These topological excitations have gained great interest as they have the potential to realize dissipationless 
spintronic/magnonic devices \cite{Li-C-Y-review-21}.

However, the validity of the conclusion concerning the topological chiral edge current of bosonic particles should be scrutinized because several important issues remain unresolved.
For instance, it has been shown theoretically that in frustrated ferromagnets,  
finite cubic interactions among bosons may yield non-perturbative damping of magnon modes 
even at $T$ = 0 \cite{PhysRevLett.117.187203}.   
This can dramatically alter the existing picture based on stable bosonic quasiparticles with topologically non-trivial properties, 
in contrast to the fermionic counterpart  where particle-number conservation forbids such cubic interactions from appearing.  
In addition, quantitative comparison of the observed thermal Hall effect with theory is difficult and often impossible 
in most frustrated quantum magnets,
because their ground states and low-energy excitations are not fully understood. 
Furthermore, recent studies on the insulating magnets such as kagome antiferromagnets with weak lattice-spin coupling
revealed the presence of non-negligible contribution of thermal Hall effect 
of phononic origin \cite{PhysRevLett.95.155901,PhysRevLett.118.145902,ideue2017giant,akazawa2020thermal}
which makes it difficult to single out the topological part. 
In fact, the thermal Hall conductivity $\kappa_{xy}$ observed in frustrated pyrochlore ferromagnet Lu$_2$V$_2$O$_7$, 
which contains a two-dimensional (2D) kagome lattice, is much larger than that expected from the DM interaction determined  
by the density functional theory \cite{PhysRevB.83.174402,PhysRevB.94.014410},
indicating that the observed $\kappa_{xy}$ may not be solely of the topological origin.   
Given these facts, it is safe to say that the topological chiral edge current of bosonic spin excitations, including its presence, remains largely unexplored.
To investigate the elusive nontrivial bosonic topology, the knowledge of the thermal Hall effect of topological bosons in magnetic systems with exactly solvable ground states, in which precise comparison between theory and experiments is possible, is crucially important.

\begin{figure}
\includegraphics[clip,width=0.8\linewidth]{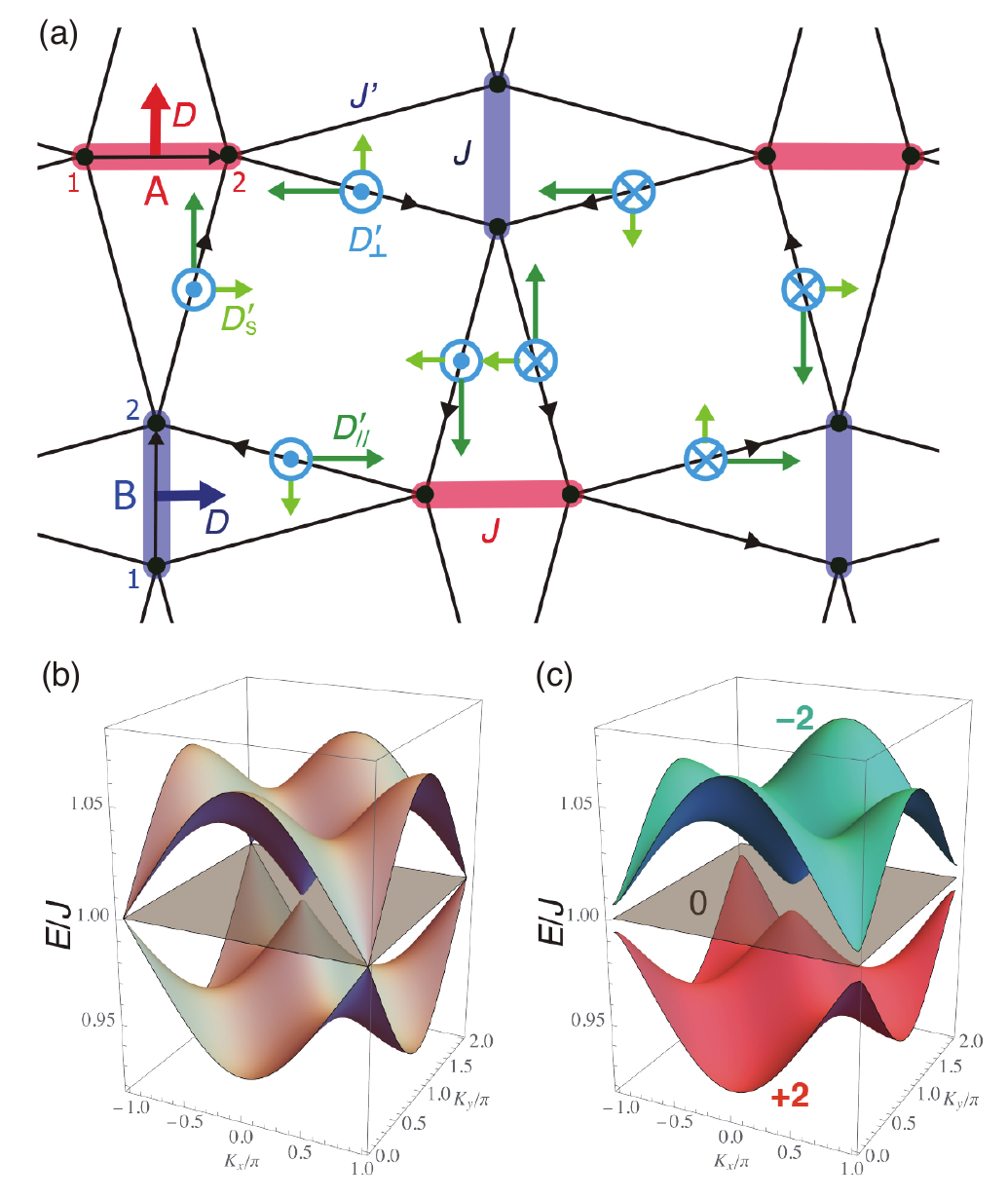}
\caption{Basic properties of $\SCBO$. (a) Shastry-Sutherland lattice and DM interactions 
in $\SCBO$
The thick red and blue lines represent the two species of dimers (A and B) of spin 1/2 Cu$^{2+}$ ions.
The black arrows ($i \to j$) indicate the order of the spins $\mathbf{S}_i$ and $\mathbf{S}_j$ in the DM interactions 
$\bolD_{ij} {\cdot} (\mathbf{S}_i \times \mathbf{S}_j)$ on the bond $(i,j)$.
The intradimer DM interaction has only an in-plane component (red and blue arrows).
For interdimer interactions, all three components are allowed by symmetry.  
Triplon bands of $\SCBO$ at zero (b) and small magnetic field (c). 
The triplon bands at zero field form a Dirac cone with three bands (two dispersive and one flat bands) touching at the $M$ point.
A small magnetic field opens the band gap, yielding the two separated bands with nontrivial Chern numbers $\pm 2$ as well as 
a trivial one. 
\label{fig:SS-lattice-and-cones}
}
\end{figure}

A candidate material that appears to be most suitable for such an investigation is $\SCBO$ \cite{PhysRevLett.82.3168}, 
a frustrated layered quantum spin system.
Each layer of this material consists of a 2D network of orthogonal dimers of spin 1/2 Cu$^{2+}$ ions 
[see Fig.\,\ref{fig:SS-lattice-and-cones}(a)].
The magnetic properties are well described by the 2D Heisenberg model with antiferromagnetic (AFM) nearest-neighbor (intradimer) and AFM next-nearest-neighbor (interdimer) exchange couplings, $J$ and $J'$, known as the Shastry-Sutherland model \cite{shastry1981exact}.  
It has been shown that for $J'/J \leq 0.675$, an unentangled product of dimer singlets is the exact ground state \cite{Corboz-M-13},  
and $\SCBO$ with $J^{\prime}/J\approx 0.63$ is believed to be in this phase \cite{miyahara2003theory}.   

The first excited state, separated from the ground state by a finite spin gap of $\sim$ 3 meV, is a bosonic $S$ = 1 quasiparticle 
dubbed {\em triplon}, 
which corresponds to exciting one of the dimer singlets to triplet \cite{PhysRevLett.84.5876}.
Without any additional interactions, the three triplon bands are triply degenerate and have very tiny dispersion of 
the order $(J'/J)^6 \sim 6\times10^{-2}$ \cite{Miyahara-U-99}.
Note that this extremely small band width is attributed not to the smallness of the inter-dimer interaction $J^{\prime}$ 
but to the unique orthogonal-dimer structure [see Fig.~\ref{fig:SS-lattice-and-cones}(a)], and the triplons in 
$\SCBO$ are essentially strongly interacting as is exemplified by, e.g., its magnetic bound states and 
the unique magnetization process \cite{miyahara2003theory,Takigawa-W-H-B-10}.     
%%%%%%%%%%%%%%%%%%%%%%%%%%%%%%%%%%%%%%%%%%%%%%%%%%%%%%%%%%%%
%%%%%%%%%%%%%%%%%%%%%%%%%%%%%%%%%%%%%%%%%%%%%%%%%%%%%%%%%%%%

Below a structural transition temperature $T_{\text{s}} \sim 395$ K, the dimer plane is slightly buckled.
As a result, DM interactions are allowed by symmetry both on each dimer and between neighboring dimers \cite{PhysRevB.75.144422,PhysRevB.83.024413}, as illustrated in Fig.\,1(a).
The known experimental values are: 
$J=3.08 \text{ meV}$, $D^{\prime}_{\perp}=-0.097 \text{ meV}$ \cite{mcclarty2017topological}.    
In the presence of the small but finite DM interactions, the triply degenerate and dispersionless triplon band is modified, 
resulting in three sub-bands with weak dispersion (of the order of the DM interactions)  
as observed in inelastic neutron scattering \cite{PhysRevLett.93.267202,PhysRevLett.113.067201,mcclarty2017topological} and electron spin resonance experiments \cite{nojiri2003esr}.

It has been proposed theoretically that the perpendicular component $D'_{\perp}$ of the interdimer DM interactions 
endows the triplon bands with a topological character \cite{romhanyi2015hall,mcclarty2017topological,PhysRevB.95.195137,PhysRevB.103.L140404} and 
turns the system into a magnetic analogue of the Chern insulators.  
Without the external magnetic field perpendicular to the dimer plane, 
the triplons form a Dirac cone with three bands touching at a single point ($M$-point) as seen in Fig.\,\ref{fig:SS-lattice-and-cones}(b).
A small magnetic field opens a band gap at the Dirac point, leading to two topological bands with the Chern numbers $\pm$2 
[Fig.\,\ref{fig:SS-lattice-and-cones}(c)].
As a result, the stable triplons moving under the momentum-space Berry flux, which is determined by the triplon band structure, are 
expected to show the thermal Hall effect.
As will be discussed later, the geometrically-suppressed kinetic energy of triplons in $\SCBO$ 
prevents their spontaneous decay at $T$ = 0 \cite{RevModPhys.85.219,PhysRevLett.117.187203} 
that may potentially invalidate the picture based on topologically non-trivial quasi-particles.
Therefore, this system seems to provide a unique playground to study the topological properties of bosonic quasiparticles.

Here we report high resolution thermal Hall conductivity $\kappa_{xy}$ measurements on the exactly solvable quantum magnet $\SCBO$.
In contrast to the theoretical predictions, we observed no discernible $\kappa_{xy}$ within our experimental resolution. 
The strong suppression of $\kappa_{xy}$ implies topological transport phenomena are strongly influenced by the inter-particle interactions in bosonic systems.

\section{Method}
Single crystals of $\SCBO$ were grown by a traveling solvent floating zone method.
The crystal was polished into a plate-like shape of roughly 1.5 mm $\times$ 0.46 mm $\times$ 0.025 mm for thermal transport measurements.
Four gold wires were attached by silver paste to serve heat links to a 1-$\mathrm{k\Omega}$ chip as a heater and three Cernox (CX1070) thermometers as seen in the inset to Fig.\,\ref{fig:kappa_xx-zero-field}.
One end of the crystal was glued to a LiF heat bath using non-metallic grease.
Thermal and thermal Hall conductivities were measured simultaneously by a standard steady-state method by applying temperature gradient $j$ and magnetic field $H$ along the crystallographic a and c axis, respectively.
To exclude the contribution from misalignment of the contacts, anti-symmetric components of the measured $\nabla_yT$ were numerically calculated and used to obtain $\kappa_{xy}(H)$.

%%%%%%%%%%%%%%%%%%%%%%%%%%%%%%%%%%%%%%%%%%%%%%%%%%%%%%%%%%%%
\section{Results}
%%%%%%%%%%%%%%%%%%%%%%%%%%%%%%%%%%%%%%%%%%%%%%%%%%%%%%%%%%%%
\begin{figure}
\includegraphics[clip,width=7cm]{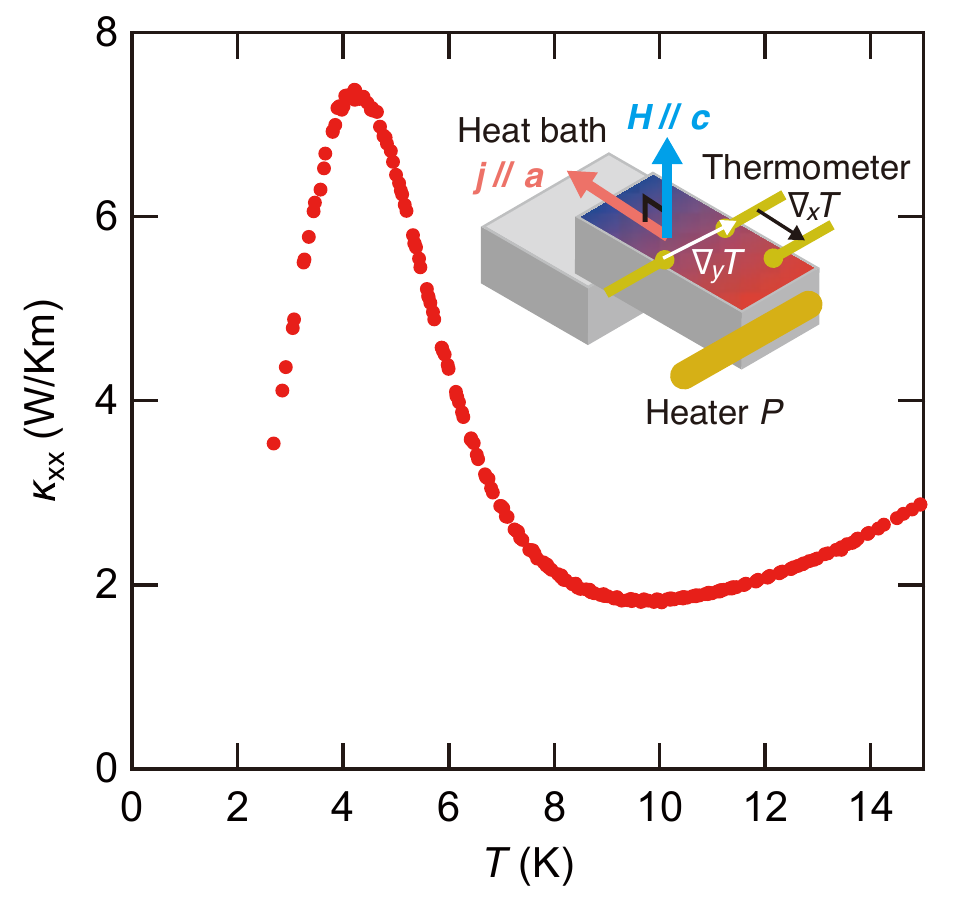}
\caption{Temperature dependence of thermal conductivity $\kappa_{xx}$ of $\SCBO$ at zero field.
The inset shows the experimental setup for thermal and thermal Hall conductivities measurements. 
\label{fig:kappa_xx-zero-field}}
\end{figure}
%%%%%%%%%%%%%%%%%%%%%%%%%%%%%%%%%%%%%%%%%%%%%%%%%%%%%%%%%%%%%%
\begin{figure*}
\includegraphics[clip,width=14cm]{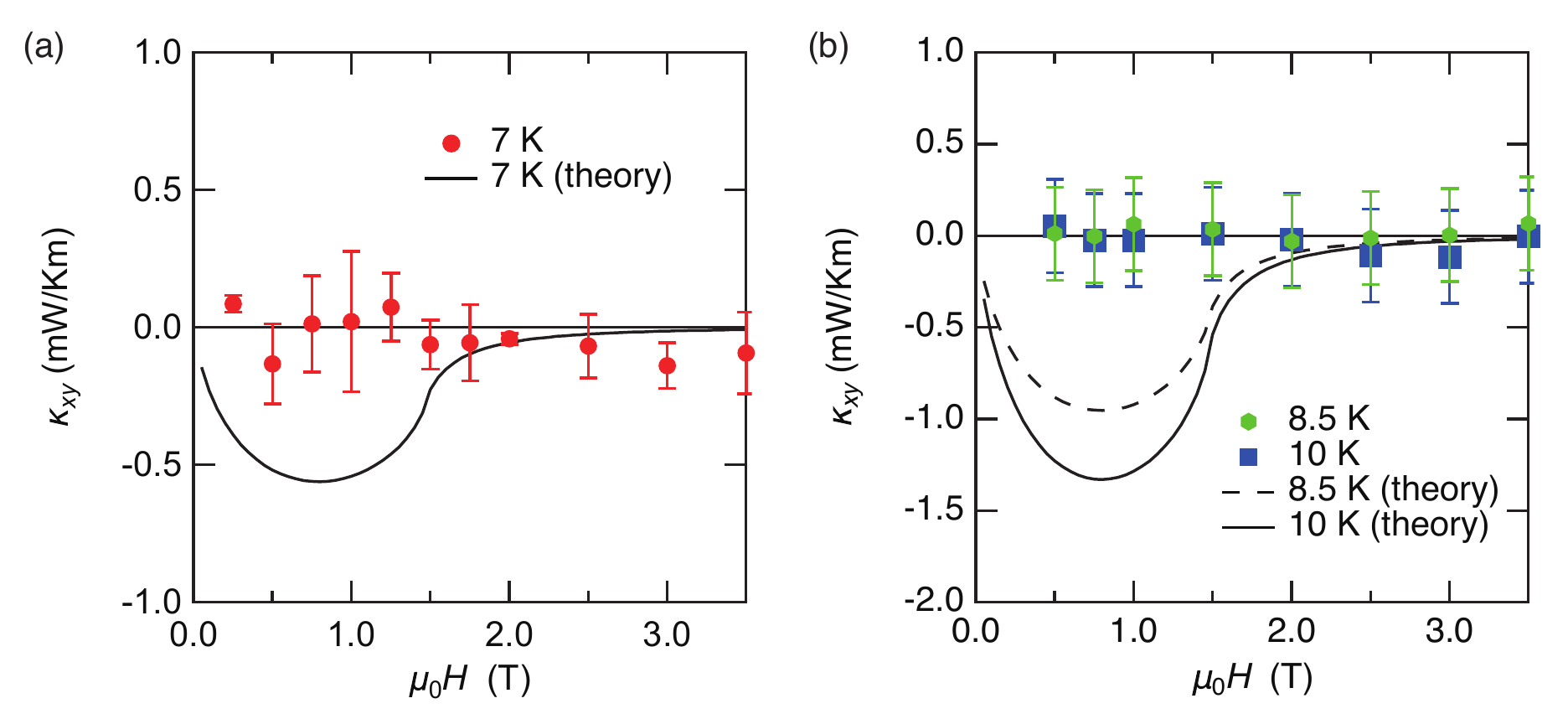}
\caption{Field dependence of thermal Hall conductivity $\kappa_{xy}$ of $\SCBO$.
(a and b) The observed $\kappa_{xy}$ at 7, 8.5, and 10\,K (red, green, and blue markers),
and the calculated $\kappa_{xy}$ at 7, 8.5 and 10\,K (solid and dashed lines).
No discernible $\kappa_{xy}$ was detected within our experimental resolution.
The error bars in the 7 K data represent a standard statistical error (see Appendix \ref{sec:experimental} for the details).
The error bars for 8.5 and 10\,K data were estimated from the largest error bar of 7 K data. 
\label{fig:kappa_xy-comparison}
}
\end{figure*}
%%%%%%%%%%%%%%%%%%%%%%%%%%%%%%%%%%%%%%%%%%%%%%%%%%%%%%%%%%%%%%

Figure\,\ref{fig:kappa_xx-zero-field} depicts the temperature-dependence of longitudinal thermal conductivity $\kappa_{xx}$ in zero field.
As the temperature is lowered, $\kappa_{xx}$ decreases gradually and then increases below 10\,K forming a peak at around 4\,K.
The magnitude of $\kappa_{xx}$ at the peak in the present crystal, which reflects the quality of the crystal, is close to the highest value 
reported so far \cite{PhysRevLett.87.047202}, indicating the high quality of our crystal.

% \subsection{Thermal Hall conductivity}
We first discuss the behaviors of the thermal Hall effect predicted theoretically.  
The solid and dashed lines in Fig.\,\ref{fig:kappa_xy-comparison}(a) and \ref{fig:kappa_xy-comparison}(b) 
represent the field-dependence of $\kappa_{xy}$ calculated for $T$ = 7, 8.5 and 10\,K  
by the formula Eq.~\eqref{eqn:triplon-kappaXY} \footnote{%
We found a factor $1/(2\pi)$ missing in the formula given in Ref.~\cite{romhanyi2015hall}, which had been probably overlooked 
in recovering the physical constants ($h$, $k_{\text{B}}$, etc.) in the formula Eq.~(35) of Supplementary Notes. We used 
the corrected formula \eqref{eqn:triplon-kappaXY} for the theoretical values.} 
which assumes stable non-interacting triplon excitations subject to the Berry curvature 
(see Appendix \ref{sec:triplon} and \ref{sec:thermal-Hall-triplon} for more information).  
We note that the topological thermal Hall effect is absent at zero temperature and appears only at finite temperatures where the triplon bands 
with nontrivial Chern numbers are thermally populated.
Its magnitude depends not only on the Berry curvature $F^{(m)}_{xy}(\bolK)$ of the topological bands but also on how these bands are occupied by 
thermally excited triplons \cite{PhysRevLett.104.066403,PhysRevLett.106.197202}.   
Finite $\kappa_{xy}$, which is negative in sign, appears at weak magnetic fields.
At around $\mu_0 H\approx 0.7$\,T, the absolute value of $\kappa_{xy}$ shows a maximum and then decreases with $H$.
Above a threshold field of $\mu_0 H_{\text{th}} \approx$ 1.5\,T, which is set by the out-of-plane component $D'_\perp$,
the Chern number changes from $\pm 2$ to zero (see Appendix \ref{sec:triplon-topology}) and the triplon bands lose their topological character, 
resulting in a rapid decrease of $\kappa_{xy}$.
Unlike the Hall conductance in fermionic band insulators where the topological bands are occupied homogeneously 
in $k$-space at $T=0$,   
the thermal Hall conductivity is not quantized in the present bosonic system;   
being the integral of the local Berry curvature and the thermal occupation factor [see Eq.~\eqref{eqn:triplon-kappaXY}], 
its value depends crucially on the detailed structure of the quasi-particle bands and is not at all universal.  
This is why the values of $\kappa_{xy}$ are very different in magnitude depending on the systems \cite{onose2010observation,watanabe2016emergence}.
In fact, even after the Chern number vanishes, small but finite $\kappa_{xy}$ survives due to the non-vanishing 
(local) Berry curvature.

In stark contrast to the theoretical predictions \cite{romhanyi2015hall}, 
however, no discernible thermal Hall signal is detected within our experimental resolution, as shown by the red circles 
in Fig.\,\ref{fig:kappa_xy-comparison}(a), and the green hexagons and the blue squares in Fig.\,\ref{fig:kappa_xy-comparison}(b).
The observed $\kappa_{xy}$ at $T$ = 7, 8.5 and 10\,K is at most 20-30\% of the theoretical values, if present. 
We also checked that this substantial discrepancy between the theoretical and experimental values is not accidental \footnote{%
At this level of approximation, a small shift of the inter-dimer exchange $J^{\prime}$ just modifies the value of the spin gap $J$ slightly, which 
may be absorbed into a small change of the temperature $T$. By directly evaluating \eqref{eqn:triplon-kappaXY}, we found that 
small changes in the DM interactions mainly shift the band-touching field but do not reduce $\kappa_{xy}$ significantly.}.

Prior to our results, it has also been reported \cite{cairns2020thermal} that $\kappa_{xy}$ is smaller than the theoretical prediction.
However, due to the large error bars of the measurements, there has been no conclusive data for the magnitude of $\kappa_{xy}$ so far.
Here our ability to control the temperature very precisely, $\delta T\sim 50\,\mu$K ($\delta T/T\sim10^{-5}$; see Appendix \ref{sec:experimental}), 
enables us to achieve high precision measurements of $\kappa_{xy}$, 
whose resolution is significantly improved from the previous report.

%%%%%%%%%%%%%%%%%%%%%%%%%%%%%%%%%%%%%%%%%%%%%%%%%%%%%%%%%%
\section{Discussion}
%%%%%%%%%%%%%%%%%%%%%%%%%%%%%%%%%%%%%%%%%%%%%%%%%%%%%%%%%%
The present results lead us to conclude that the thermal Hall effect of topological origin predicted theoretically \cite{romhanyi2015hall} is   
absent in $\SCBO$.   
In this section, we examine several possible reasons for this large discrepancy between the theoretical prediction based on the picture 
of stable non-interacting quasi-particles and the experimental results.   
%%%%%%%%%%%%%%%%%%%%%%%%%%%%%%%%%%%%%%%%%%%%%%%%%%%%%%%%%%
\subsection{Domain formation}
%%%%%%%%%%%%%%%%%%%%%%%%%%%%%%%%%%%%%%%%%%%%%%%%%%%%%%%%%%
We begin by considering a few extrinsic sources of the strong suppression of $\kappa_{xy}$.  
First of all, the sample is expected to contain structural domains with different patterns of buckling in the ab-plane.  
The buckling is crucial for the non-zero thermal Hall response as it induces the in-plane component $D^{\prime}_{\parallel}$ 
of the DM interactions \cite{PhysRevB.83.024413} which is necessary for the finite Berry curvature 
[see Eq.~\eqref{eqn:triplon-Berry-curvature}].  
However, the absence of the thermal Hall effect cannot be attributed to the domain formation for the following reason. 
To be specific, we assume that the A (B) dimers are shifted upward (downward) from the basal planes.  
Then, in a system where the directions of the shift are reversed, all the {\em in-plane} components of 
the DM interactions (i.e., $D$, $D^{\prime}_{\text{s}}$, and $D^{\prime}_{\parallel}$) are flipped.  However, 
since both the triplon dispersion \eqref{eqn:triplon-dispersion} and the Berry curvature \eqref{eqn:triplon-Berry-curvature} 
depend only on ${D^{\prime}_{\parallel}}^{2}$ \footnote{%
Precisely, these quantities depend on its effective value },  
$\kappa_{xy}$ from different domains have the same sign and do not cancel each other.

%%%%%%%%%%%%%%%%%%%%%%%%%%%%%%%%%%%%%%%%%%%%%%%%%%%%%%%%%%
\subsection{Effects of inter-layer couplings}
\label{sec:inter-layer}
%%%%%%%%%%%%%%%%%%%%%%%%%%%%%%%%%%%%%%%%%%%%%%%%%%%%%%%%%%
The interactions among the Shastry-Sutherland layers may affect the thermal Hall conductivity \eqref{eqn:triplon-kappaXY} 
calculated for the purely two-dimensional system. 
 In the actual $\SCBO$ compound, there are several kinds of interlayer interactions that might modify 
 the 2D physics.  First of all, there are exchange couplings of the form on the shortest bonds  connecting the adjacent layers \cite{miyahara2003theory}:
 \begin{equation}
J_{\perp} \left( \bolS_{\text{A},1}+\bolS_{\text{A},2}\right) {\cdot} \left( \bolS_{\text{B},1}+\bolS_{\text{B},2}\right)
= J_{\perp}  \bolT_{\text{A}} {\cdot}  \bolT_{\text{B}}  \; ,
 \end{equation}
 where A and B dimers respectively are on the upper and lower layers, or vice versa.   
 As this interaction vanishes when at least one of A and B is occupied by singlet, it does not help a singlet triplon to hop between adjacent layers
 \footnote{%
If we take into account the on-dimer DM-inteaction $D$, $\bolT_{\text{A,B}}$ contain the order $D$ terms of the form 
$\left( \tilde{s}_{\text{A},\text{B}}^{\dagger} \tilde{t}_{\text{A},\text{B},a} 
+ \tilde{t}^{\dagger}_{\text{A},\text{B},a}\tilde{s}_{\text{A},\text{B}} \right)$, and $J_{\perp}$ can generate inter-layer triplon hopping 
processes (between A and B dimers) 
proportional to $D^{2}J_{\perp}/ J^{2}$.  As these processes are much smaller than the inter-layer hopping due to 
the inter-layer DM interactions, we may neglect them at this level of approximation.}.  
This conclusion remains the same even when many triplons exist. 
If we note that the $J_{\perp}$ interaction itself does not change the positions of triplons, we see that the inter-layer interaction 
$J_{\perp}$ does not contribute to any kinds of trilpon motion at all.

\begin{figure}[htb]
\begin{center}
\includegraphics[scale=0.3]{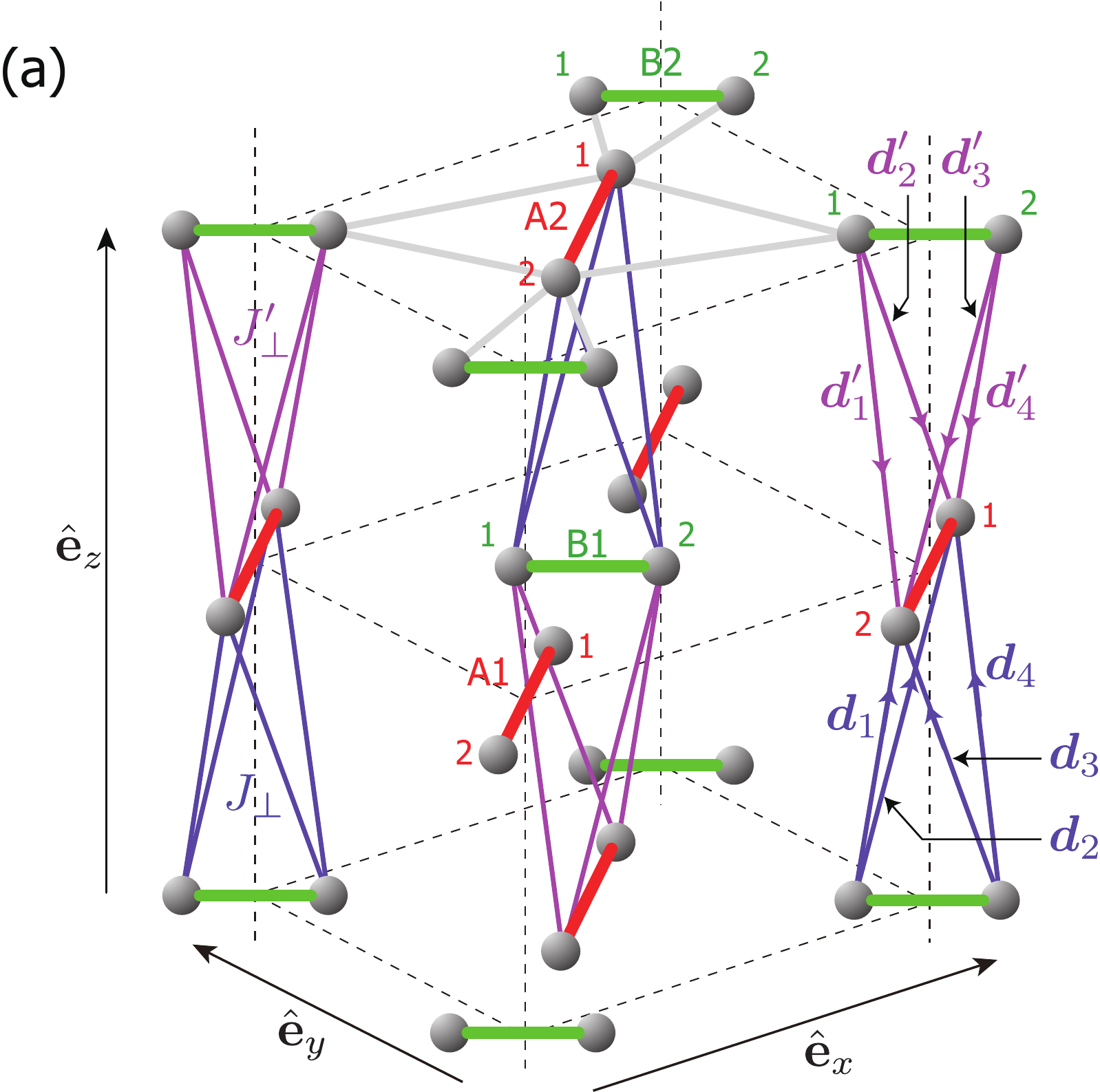}
\vspace{5mm}

\includegraphics[scale=0.4]{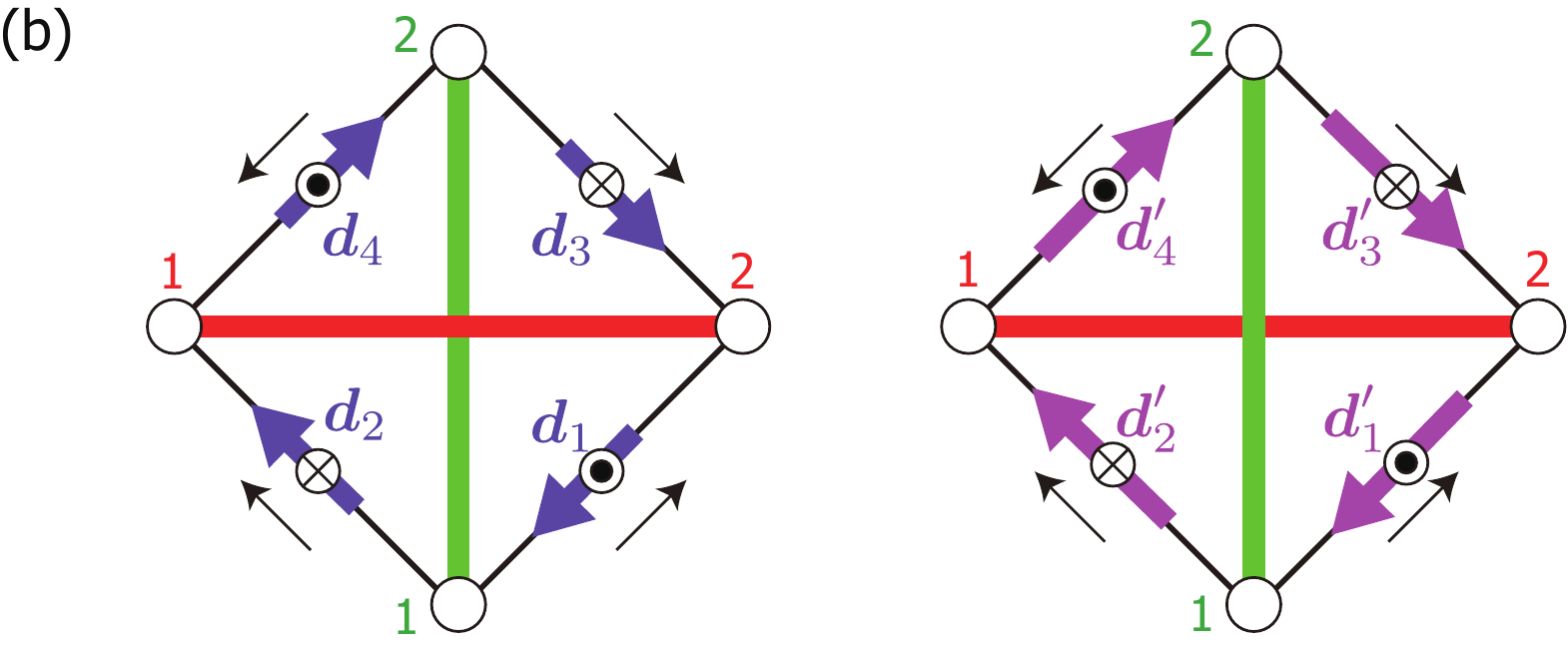}
\end{center}
\caption{3D structure of $\SCBO$ and symmetry-allowed interlayer DM interactions (a), and 
its top view (b). 
Due to the buckling in $ab$-plane, the interlayer DM interactions can be  
different for A-B (left; $d_{y,z}$) and B-A (right; $d_{y,z}^{\prime}$) pairs.  
\label{fig:interlayer-DM}}
\end{figure}

\begin{figure}[htb]
\begin{center}
\includegraphics[width=\columnwidth,clip]{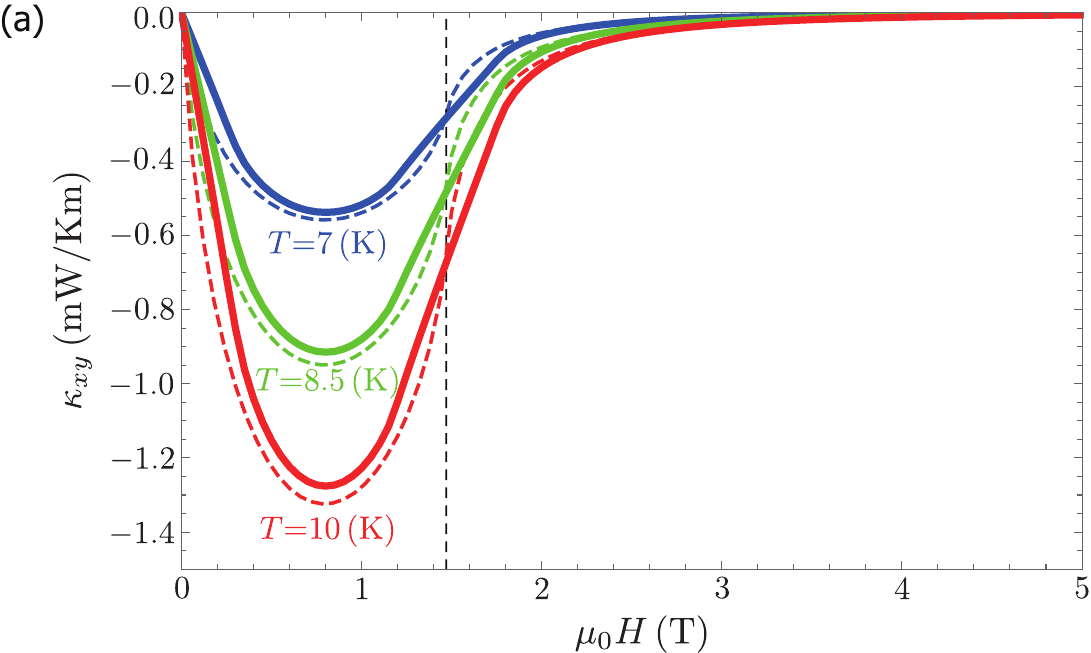}
\vspace{5mm}

\includegraphics[width=\columnwidth,clip]{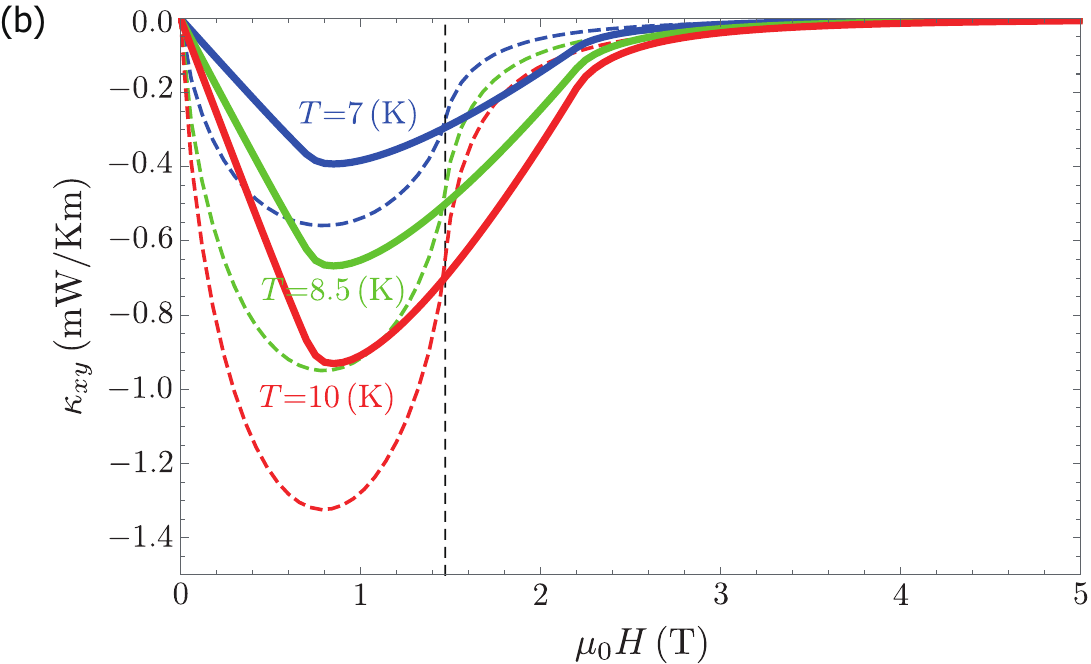}
\end{center}
\caption{Thermal Hall conductivity $\kappa^{\text{3D}}_{xy} (T,H_z)$ (solid lines) and 
its 2D limit $\kappa_{xy} (T,H_z)$ (dashed lines) for different temperatures.  
The interlayer DM interactions are assumed to be (a) $d_z = 0.2 D^{\prime}_{\perp}$ or 
(b) $d_z = 0.5 D^{\prime}_{\perp}$. At $\mu_{0} H=1.47$ T (dashed line), a band-touching transition occurs and the triplon bands lose their topological character. 
\label{fig:kappa_xy-3D}}
\end{figure}

\begin{figure}[htb]
\begin{center}
\includegraphics[scale=0.4]{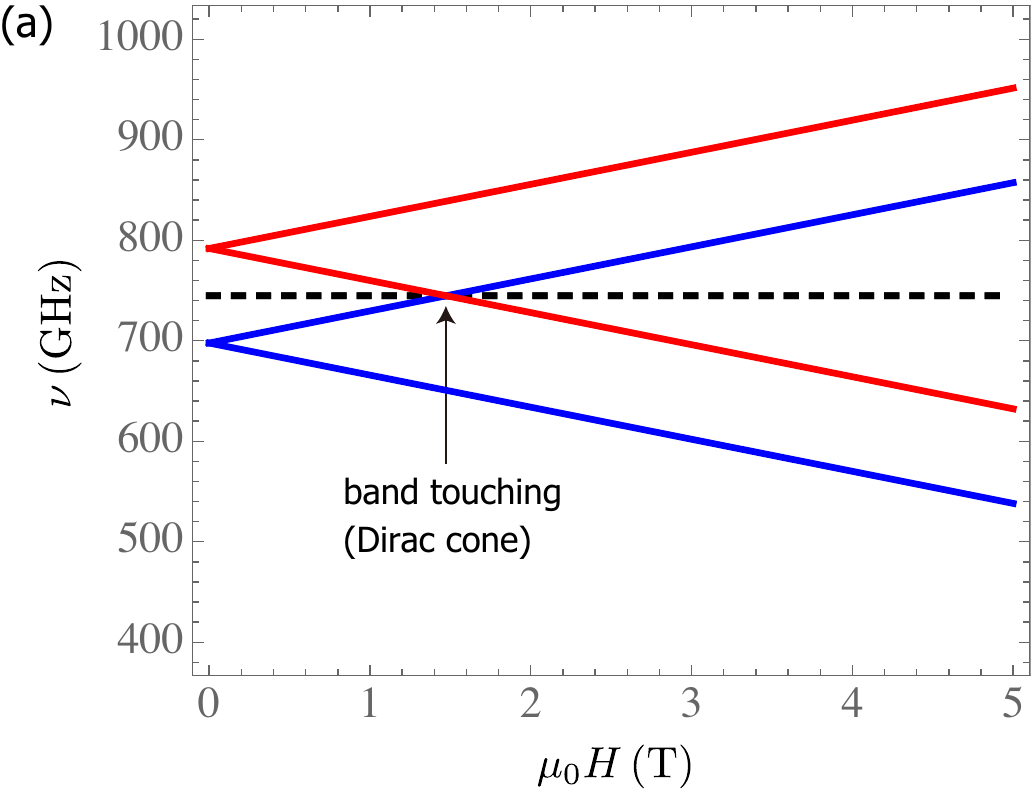}
\vspace{5mm}

\includegraphics[scale=0.4]{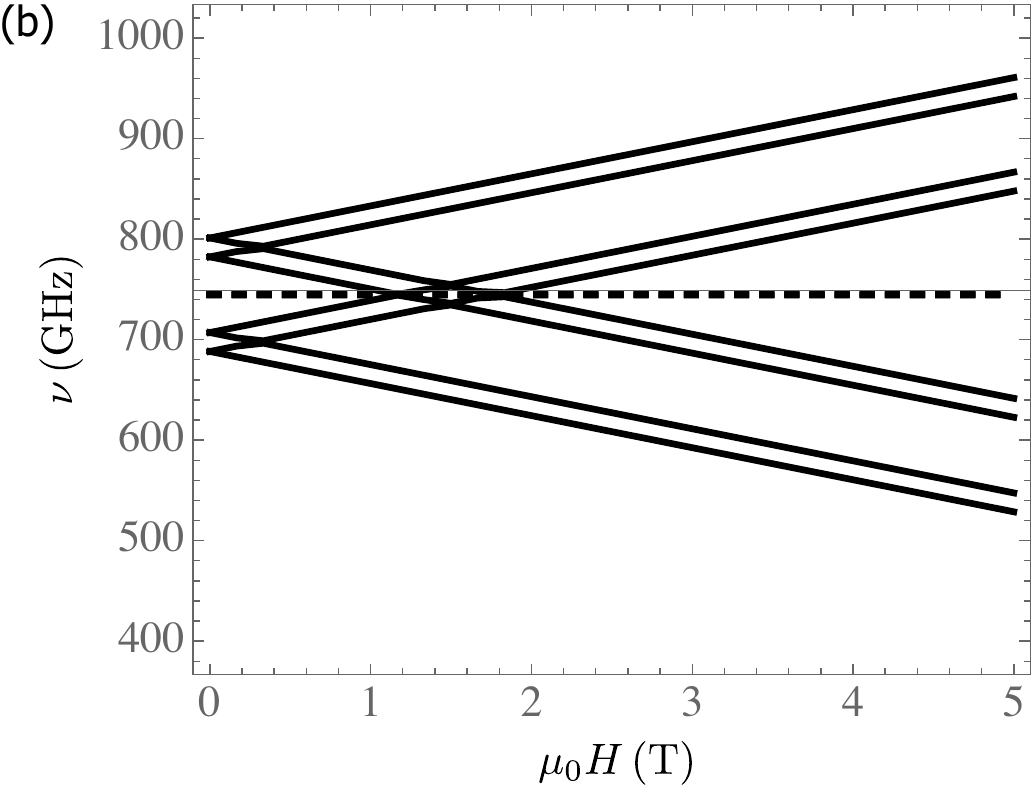}
\includegraphics[scale=0.4]{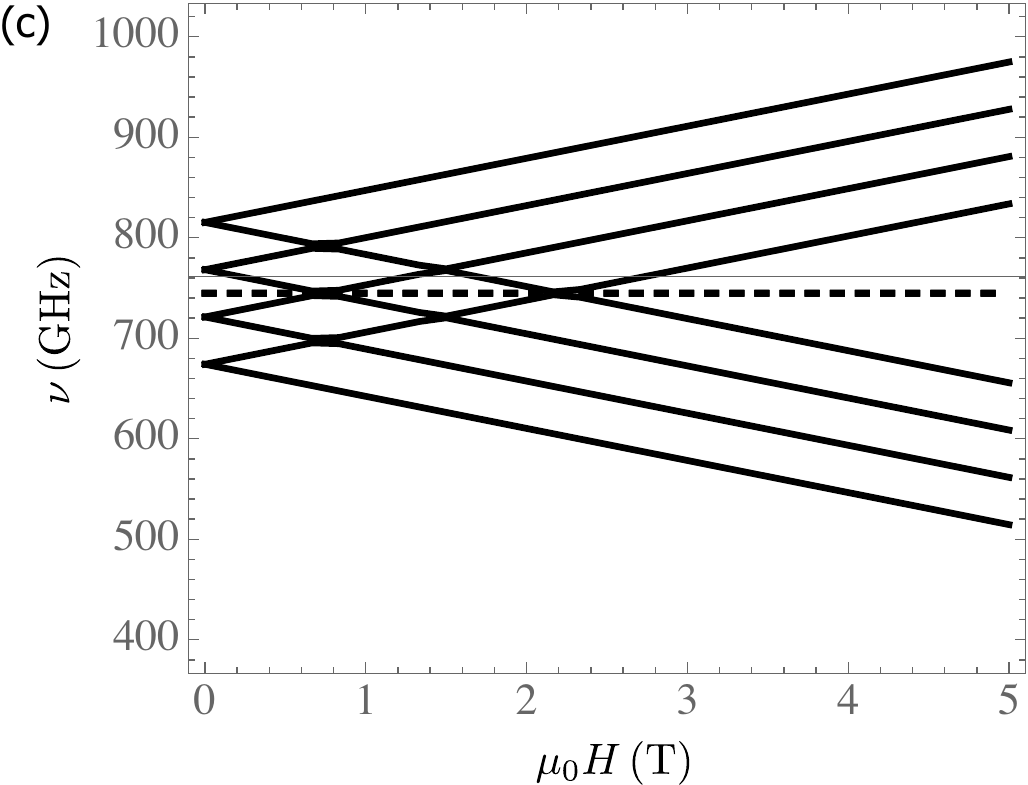}
\end{center}
\caption{Energy levels at $\Gamma$-point which are to be compared with ESR spectrum in, e.g., Ref.~\cite{nojiri2003esr}: 
for (a) $d_{z}=0$, (b) $d_{z}=0.2 D^{\prime}_{\perp}$, and (c) $d_{z}=0.5 D^{\prime}_{\perp}$.  
At the level crossing at $\mu_{0} H=1.47$ T (left panel), the triplon bands lose their topological properties.  
The six levels for $d_z =0$ are further split by interlayer DM interactions $d_z$.
\label{fig:ESR-12-band}}
\end{figure}

On top of the usual exchange couplings, the crystal symmetry allows several DM interactions between layers as shown in Fig.~\ref{fig:interlayer-DM}.  
It is also suggested \cite{Bhowmick-S-SCBO-20} that the interlayer DM interactions allowed by symmetry 
can change the topological properties of an isolated plane.   
We calculated the triplon spectrum and the thermal Hall conductivity in the presence of such interlayer DM interactions:  
\begin{equation}
 \kappa_{xy}^{\text{3D}} (T,H_z)  
 = \frac{1}{\pi} \int_{H_z  - \frac{2|d_z|}{g_z \mu_{\text{B}}}}^{H_z  + \frac{2|d_z|}{g_z \mu_{\text{B}}}} \! 
 \frac{ \kappa_{xy}(T,h) }{\sqrt{\left( \frac{2d_z}{g_z \mu_{\text{B}}} \right)^{2}-(h -H_z)^{2}}} dh  
\label{eqn:hz-integral-for-kappa3D}
\end{equation}
(see Appendix \ref{sec:inter-layer-appendix} for the derivation).  
In deriving the above formula, we have assumed, for simplicity, that the $z$-components of the eight inter-layer DM interactions have 
a common absolute value $d_{z}$.  
We plot $\kappa_{xy}^{\text{3D}} (T,H_z)$ calculated by Eq.~\eqref{eqn:hz-integral-for-kappa3D}  
in Fig.~\ref{fig:kappa_xy-3D} for two different values of the inter-layer DM interaction $d_{z}$. 
No precise information is available on the values of the inter-layer DM interactions in $\SCBO$ for now, 
we can obtain a crude estimate of $d_{z}$ by comparing the theoretical spectrum with the ESR results \cite{nojiri2003esr}.  
We plot in Fig.~\ref{fig:ESR-12-band} the spectrum of the twelve bands at the $\Gamma$-point 
to find that the values of the inter-layer DM interactions $d_{z}$ which are compatible with the ESR results \cite{nojiri2003esr} do not 
reduce $\kappa_{xy}$ substantially.
%%%%%%%%%%%%%%%%%%%%%%%%%%%%% FIG %%%%%%%%%%%%%%%%%%%%%%%
%%%%%%%%%%%%%%%%%%%%%%%%%%%%%%%%%%%%%%%%%%%%%%%%%%%%%%%
%%%%%%%%%%%%%%%%%%%%%%%%%%%%% FIG %%%%%%%%%%%%%%%%%%%%%%%%%%%%%%%
%%%%%%%%%%%%%%%%%%%%%%%%%%%%%%%%%%%%%%%%%%%%%%%%%%%%%%%%%%%%%%%
%%%%%%%%%%%%%%%%%%%%%%%%%%%%% FIG %%%%%%%%%%%%%%%%%%%%%%%
%%%%%%%%%%%%%%%%%%%%%%%%%%%%%%%%%%%%%%%%%%%%%%%%%%%%%%

%%%%%%%%%%%%%%%%%%%%%%%%%%%%%%%%%%%%%%%%%%%%%%%%%%%%%%%%%%
\subsection{Magneto-phonon coupling}
%%%%%%%%%%%%%%%%%%%%%%%%%%%%%%%%%%%%%%%%%%%%%%%%%%%%%%%%%%
A magneto-phonon coupling, which has been pointed out by a resonant scattering observed in the longitudinal thermal conductivity measurements \cite{PhysRevLett.87.047202}, can be an origin of the triplon damping.
However, because the magneto-phonon coupling is most active around 20 K, such a coupling is unlikely to be an origin of the strong suppression of the thermal Hall conductivity at lower temperatures (e.g., 7-10 K). 
Indeed, a correlated decay process of the triplon excitations which is responsible for the triplon damping at $T \leq 10$\,K \cite{PhysRevLett.113.067201} cannot be explained solely by acoustic phonons.

%%%%%%%%%%%%%%%%%%%%%%%%%%%%%%%%%%%%%%%%%%%%%%%%%%%%%%%%%%%%%%%%
\begin{figure}
\includegraphics[clip,width=8.6cm]{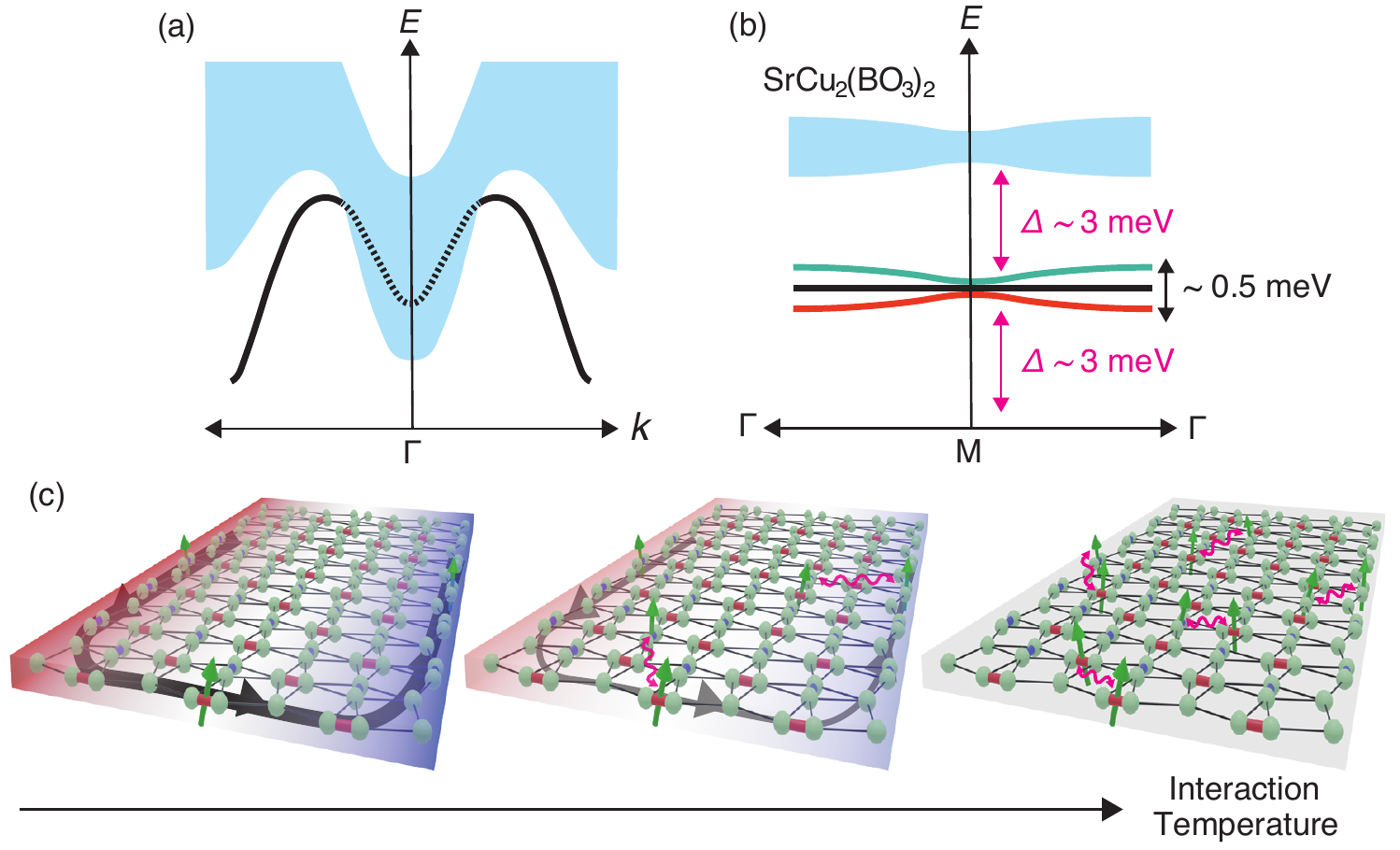}
\caption{Damping of bosonic spin excitations and destruction of topological properties.
(a) Single magnon band (solid lines) and two magnon continuum (light blue area) in $S$ = 1 Heisenberg chain.
The single magnon band overlaps the two magnon continuum around the $\mathrm{\Gamma}$ point, yielding the non-perturbative damping of magnons.
(b) Single triplon band (solid lines) and two triplon continuum (light blue area) in $\SCBO$.
The spin gap $\Delta \sim$ 3 meV, much larger than the triplon band width of $\sim$ 0.5 meV, avoids 
an overlap of the single triplon band and two-triplon continuum,
and therefore the spontaneous decay does not occur at 0 K.
(c) Schematic illustration of the triplon edge current in the presence of interactions between thermally populated triplons in $\SCBO$.
The left panel displays the triplon edge current in the absence of the interaction.
With increasing the interaction and/or temperature, the topologically nontrivial properties are severely affected by the interaction between thermally excited bosons and are eventually almost destroyed, as illustrated by the middle and right panels. 
\label{eqn:cartoon-edge}
}
\end{figure}
%%%%%%%%%%%%%%%%%%%%%%%%%%%%%%%%%%%%%%%%%%%%%%%%%%%%%%%%%%%%%%%%
%%%%%%%%%%%%%%%%%%%%%%%%%%%%%%%%%%%%%%%%%%%%%%%%%%%%%%%%%%
\subsection{Triplon-triplon interaction}
%%%%%%%%%%%%%%%%%%%%%%%%%%%%%%%%%%%%%%%%%%%%%%%%%%%%%%%%%%
Having excluded the extrinsic sources of the suppression, 
one may ask about the stability of triplon quasi-particles against additional interactions which have not been 
included in the previous analyses \cite{romhanyi2015hall,mcclarty2017topological}.
The realization of topological magnon bands in other spin systems has been proposed for the frustrated kagome ferromagnets, 
such as $\text{Lu$_2$V$_2$O$_7$}$ \cite{onose2010observation,PhysRevB.85.134411} and Cu[1,3-benzenedicarboxylate(bdc)] \cite{PhysRevLett.115.147201}.
It has been shown theoretically that in these systems the DM interactions not only provide 
the quasi-particles with non-trivial topological characters, 
but also generate anharmonic (e.g., cubic) interactions among them leading to their damping \cite{PhysRevLett.117.187203}.
To be specific, we here consider the anharmonic cubic coupling of the form $b^\dag b^\dag b$ 
[with  $b^\dag$ ($b$) being the bosonic creation (annihilation) operator for triplons or magnons], 
which has been discussed as an origin of the magnon damping \cite{RevModPhys.85.219}.
Such a cubic interaction leads to the decay of a single bosonic quasi-particle (with energy $\epsilon_\mathbf{k} $) into two satisfying 
the conservation of energy and momentum: $\epsilon_\mathbf{k} = \epsilon_\mathbf{q} + \epsilon_{\mathbf{k}-\mathbf{q}}$.  
For this constraint to be satisfied in the decay process, an overlap between the two-particle continuum and a single-particle 
band is required, as illustrated in Fig.\,\ref{eqn:cartoon-edge}(a).
In the above kagome ferromagnets with such an overlap, cubic interactions induced by DM interactions in fact
lead to strong decay of topological magnons even at $T$ = 0, as pointed out in Ref. \cite{PhysRevLett.117.187203}.

It is known that such a cubic interaction $b^\dag b^\dag b$ (of the order of $J^{\prime}$) exists in $\SCBO$ 
already in the absence of the DM interactions \cite{Totsuka-M-U-01}.  However, 
it should be stressed that, in stark contrast to the magnons with sizable dispersions of the order of the exchange interactions 
found in the above frustrated magnets, the prerequisite overlap is absent in the triplons in $\SCBO$.   
In fact, as shown in Fig.\,\ref{eqn:cartoon-edge}(b), the triplon band width ($\sim$ 0.5 meV) is substantially suppressed 
in comparison to the spin gap ($\sim$ 3 meV) due to the unique (i.e., orthogonal-dimer) lattice geometry  
of $\SCBO$   
and the two-triplon continuum is well separated from the single-triplon band \cite{PhysRevLett.93.267202}.   
Therefore, this decay process is forbidden at $T$ = 0, suggesting that the damping caused by the anharmonic couplings is expected to be much smaller if present. 

Therefore, we need to consider the effects of finite temperatures and 
the interaction among the thermally excited triplons, which has not been taken into account seriously in the previous calculations.
One plausible origin for the strong suppression of $\kappa_{xy}$ may be triplon damping at finite temperatures found 
in inelastic neutron scattering measurements \cite{PhysRevLett.113.067201}.  
However, the strong suppression of $\kappa_{xy}$ cannot be simply understood by the quasi-particle damping alone 
for the following reason.    
Although the experiments observed a strong finite-temperature damping of the triplon excitations,  
the well-defined peak associated with a single-triplon excitation still remains below 10\,K \cite{PhysRevLett.113.067201}.
In particular, at $T$ = 7\,K, at which the thermal Hall measurements were done, the sharp single-triplon peak is observed, 
which means that a {\em single} triplon quasi-particle is stable at low enough temperatues.
Therefore, as long as the picture based on {\it non-interacting} quasi-particles of topological characters  
is correct, it is unlikely that the band broadening due to the damping has such a dramatic effect as to wipe out 
the thermal Hall signal almost completely. 
In this respect, the absence of $\kappa_{xy}$ is surprising, implying that, in contrast to fermionic systems in which non-interacting limit 
is well-defined, topological transport phenomena in their bosonic counterparts are substantially influenced 
by small but finite particle-particle interactions.

%%%%%%%%%%%%%%%%%%%%%%%%%%%%%%%%%%%%%%%%%%%%%%%%%%%%%%%%%%%%%%
\section{Summary and outlook}
%%%%%%%%%%%%%%%%%%%%%%%%%%%%%%%%%%%%%%%%%%%%%%%%%%%%%%%%%%%%%%
In summary, we measured the thermal Hall conductivity of the exactly solvable quantum magnet $\SCBO$ and compared the results 
with the theoretical values calculated by assuming the existence of stable non-interacting quasi-particles (triplons).  
According to our calculations, the sign of $\kappa_{xy}$ is negative, and 
the magnitude is reduced by a factor of 2$\pi$, as opposed to the previous theoretical predictions.  
The measurements were performed with extremely high accuracy, and   
we observed no discernible $\kappa_{xy}$, which is at most 20-30\% of our calculations if present. 
The strong suppression of $\kappa_{xy}$ cannot be simply explained by the triplon damping at finite temperature, 
indicating that the inter-particle interactions dramatically alter the topological transport properties.
These arguments suggest that, in order to understand the thermal Hall transport in $\SCBO$ 
even qualitatively, it is crucially important to take into account the strong interactions among a macroscopic number of thermally-excited triplons [see Fig.\,\ref{eqn:cartoon-edge}(c)], which requires further theoretical investigations.
Higher-order interactions between bosonic particles, which have never been considered in the topological context, may be an origin for the disappearance of the topological thermal Hall effect.
The present study also calls special attention to the existing interpretation of the thermal Hall effect observed 
in various insulating magnets in terms of topological bosonic excitations. 

\begin{acknowledgments}
We thank Karlo Penc, Judit Romhanyi, and Mike Zhitomirsky for insightful discussions.
This work is supported by Grants-in-Aid for Scientific Research (KAKENHI) (Nos. JP18H01177, JP18H01180, JP18H05227, JP21H04443, and 21K13881)
and on Innovative Areas ``Quantum Liquid Crystals'' (No. 19H05824) from the Japan Society for the Promotion of Science, and JST CREST (JPMJCR19T5).
\end{acknowledgments}
%%%%%%%%%%%%%%%%%%%%%%%%%%%%%%%%%%%%%%%%%%%%%%%%%%%%%%%%
\appendix
%%%%%%%%%%%%%%%%%%%%%%%%%%%%%%%%%%%%%%%%%%%%%%%%%%%%%%%%
% %%%%%%%%%%%%%%%%%%%%%%%%%%%%%%%%%%%%%%%%%%%%%%%%%%%%%%%%%%%%
\section{Triplons in orthogonal-dimer system}
\label{sec:triplon}
%%%%%%%%%%%%%%%%%%%%%%%%%%%%%%%%%%%%%%%%%%%%%%%%%%%%%%%%%%%%
%%%%%%%%%%%%%%%%%%%%%%%%%%%%%%%%%%%%%%%%%%%%%%%%%%%%%%%%%%%%
\subsection{Ground state and low-lying excitations}

%%%%%%%%%%%%%%%%%%%%%%%%%%%%%%%%%%%%%%%%%%%%%%%%%%%%%%%%%%%%
It is well-known that the basic properties of $\SCBO$ below a structural transition 
at $T_{\text{s}} \sim 395$ K are well described by the Shastry-Sutherland model \cite{shastry1981exact} 
with the DM interactions added \cite{PhysRevB.75.144422,PhysRevB.83.024413} [see Fig.\,1(a)]:  
\begin{equation}
\begin{split}
\mathcal{H}_{\text{SCBO}} = & 
J \sum_{\text{n.n.}} \bolS_{i}{\cdot}\bolS_{j} + J^{\prime} \sum_{\text{n.n.n.}} \bolS_{i}{\cdot}\bolS_{j} 
- \mu_{\text{B}} \sum_{i} \mathbf{H} \mathbf{g}_{i} \bolS_{i} \\
& + \sum_{\text{n.n.}} \bolD_{ij} {\cdot} ( \bolS_{i}{\times} \bolS_{j}) 
+ \sum_{\text{n.n.n.}} \bolD^{\prime}_{ij} {\cdot} ( \bolS_{i}{\times} \bolS_{j})  \; ,
\label{eqn:SCBO-model-Ham}
\end{split}
\end{equation}
where $\sum_{\text{n.n.}}$ ($\sum_{\text{n.n.n.}}$) denotes the summation over nearest-neighbor 
(next-nearest-neighbor) spin pairs $(\bolS_i,\bolS_j)$, and, to ease the notations, we have included the magnetic constant $\mu_{0}$ 
in the definition of $\mathbf{H}$.  
In the absence of the DM interactions, the exact ground state for $J'/J \leq 0.675$ is given by putting spin-singlets on all the dimer bonds.  
The low-lying excitation over the exact ground state is given by exciting one of the dimer singlets into triplet 
({\em triplon}).  
The inter-dimer exchange interaction ($J^{\prime}$) gives only an isolated triplon 
an extremely tiny dispersion $\sim (J'/J)^6 \sim 6\times10^{-2}$ \cite{Miyahara-U-99} except for renormalizing 
the the spin gap (or, the effective dimer coupling $J$).   On the other hand, the DM interactions allowed by 
crystallographic symmetry provide the triplons with a larger band width of the order $\sim D^{\prime}$.   
Therefore, as far as the (cubic and quartic) interactions can be neglected, 
it would be legitimate to consider the effective Hamiltonian in which only the hopping of the order of 
the DM interactions is kept and $J$ is replaced with its renormalized value (i.e., the observed spin gap).   
The resulting triplon Hamiltonian (the kernel of the quadratic part of the triplon Hamiltonian, precisely) 
can be written in the following suggestive form \cite{romhanyi2015hall}: 
\begin{equation}
\mathcal{H}_{\text{3-band}} (K_x,K_y) 
= J\mathbb{I} + \bolB (K_x,K_y)  {\cdot} \boldsymbol{\mathcal{T}}  \; ,
\label{eqn:3-level-model}
 \end{equation}
 with the three matrices 
 \[
\mathcal{T}_x =
\begin{pmatrix}
0 & 0 & 0 \\ 0 & 0 & -1 \\ 0 & -1 & 0 
\end{pmatrix}  , \;
\mathcal{T}_y =
\begin{pmatrix}
0 & 0 & -1 \\ 0 & 0 & 0 \\ -1 & 0 & 0 
\end{pmatrix}  , \;
\mathcal{T}_z =
\begin{pmatrix}
0 & i & 0 \\ - i & 0 & 0 \\ 0 & 0 & 0 
\end{pmatrix} 
\]
satisfying the usual spin commutation relations and the  fictitious ``magnetic field'' $\bolB$ defined by:
 \begin{equation}
 \begin{split}
 & \bolB(K_x,K_y)  \\
&=  \left( 
 \overline{D}^{\prime}_{\parallel} \sin K_y , \,  \overline{D}^{\prime}_{\parallel} \sin K_x , \, 
g_z \mu_{\text{B}} H - D^{\prime}_{\perp}  (\cos K_x +\cos K_y) 
 \right)   \;   .
 \end{split}
 \label{eqn:3-band-B-vector-1}
\end{equation}
Thanks to the simple form \eqref{eqn:3-level-model}, the three triplon bands are explicitly written down as
\begin{equation}
\epsilon^{(m)}(K_x,K_y) 
= J + m | \bolB (K_x,K_y) |  \quad   (m=-1,0,1)  \; .
\label{eqn:triplon-dispersion}
\end{equation}   
When the external field perpendicular to the dimer plane is absent ($H=0$), there are two Dirac cones with three bands touching 
at $(K_{x},K_{y})=(\pi,0)$ and $(0,\pi)$ corresponding to 
the existence of the two inequivalent dimers (A and B) in a unit cell [Fig.~\ref{fig:SS-lattice-and-cones}(b)].  
%%%%%%%%%%%%%%%%%%%%%%%%%%%%%%%%%%%%%%%%%%%%%%%%%%%%%%%%%%%%
\subsection{Topology of triplon bands}
\label{sec:triplon-topology}
%%%%%%%%%%%%%%%%%%%%%%%%%%%%%%%%%%%%%%%%%%%%%%%%%%%%%%%%%%%%
A small perpendicular magnetic field $H$ opens a band gap ($\sim H$) at the Dirac point [Fig.\,\ref{fig:SS-lattice-and-cones}(c)], 
leading to two topological bands (top and bottom) with the Chern numbers $\pm$2 and one trivial one in the middle.  
The perpendicular component $D'_{\perp}$ of the interdimer DM interactions 
endows the triplon bands with a topological character \cite{romhanyi2015hall,mcclarty2017topological,PhysRevB.95.195137,PhysRevB.103.L140404} and 
turns the system into a magnetic analogue of the Chern insulators.   
Specifically, the three split triplon bands ($m=-1,0,1$) have the following Berry curvatures in the momentum space \footnote{%
Here the momentum $(K_{x},K_{y})$ is defined with respect to an effective square lattice obtained by neglecting 
the difference between the A and B dimers. It has a Brillouin zone twice larger than that of the original Shastry-Sutherland lattice.}:
\begin{equation}
\begin{split}
& F^{(m)}_{xy}(K_x,K_y)  \\
& = 
- \frac{ m \left\{ D^{\prime}_{\perp} \left(\cos K_x +\cos K_y \right) - g_z \mu_{\text{B}} H \cos K_x \cos K_y \right\}  }%
{| \bolB (K_x,K_y) |^{3}}  \overline{D}^{\prime}_{\parallel}{}^2 \;  .
\end{split}
\label{eqn:triplon-Berry-curvature}
\end{equation}
As a result, the stable triplons moving under the $K$-space Berry curvature, which is determined by the triplon band structure, are 
expected to exhibit the thermal Hall effect \cite{PhysRevLett.104.066403,PhysRevLett.106.197202}.   
The non-trivial topology of the triplon bands can be characterized by the Chern numbers defined by:
\begin{equation}
C^{(m)} = 
\frac{1}{2\pi} \int_{\text{B.Z.}} \! d^{2}\bolK  \,  F^{(m)}_{xy}(K_x,K_y)  
\quad (m= -1,0,1)  \; ,
\label{eqn:H-vs-Chern}
\end{equation}
which, for weak enough field, take the values $C^{(m)}= -2m$ [see Fig.~\ref{fig:SS-lattice-and-cones}(c)].    

\begin{figure}[H]
\begin{center}
\includegraphics[scale=0.4]{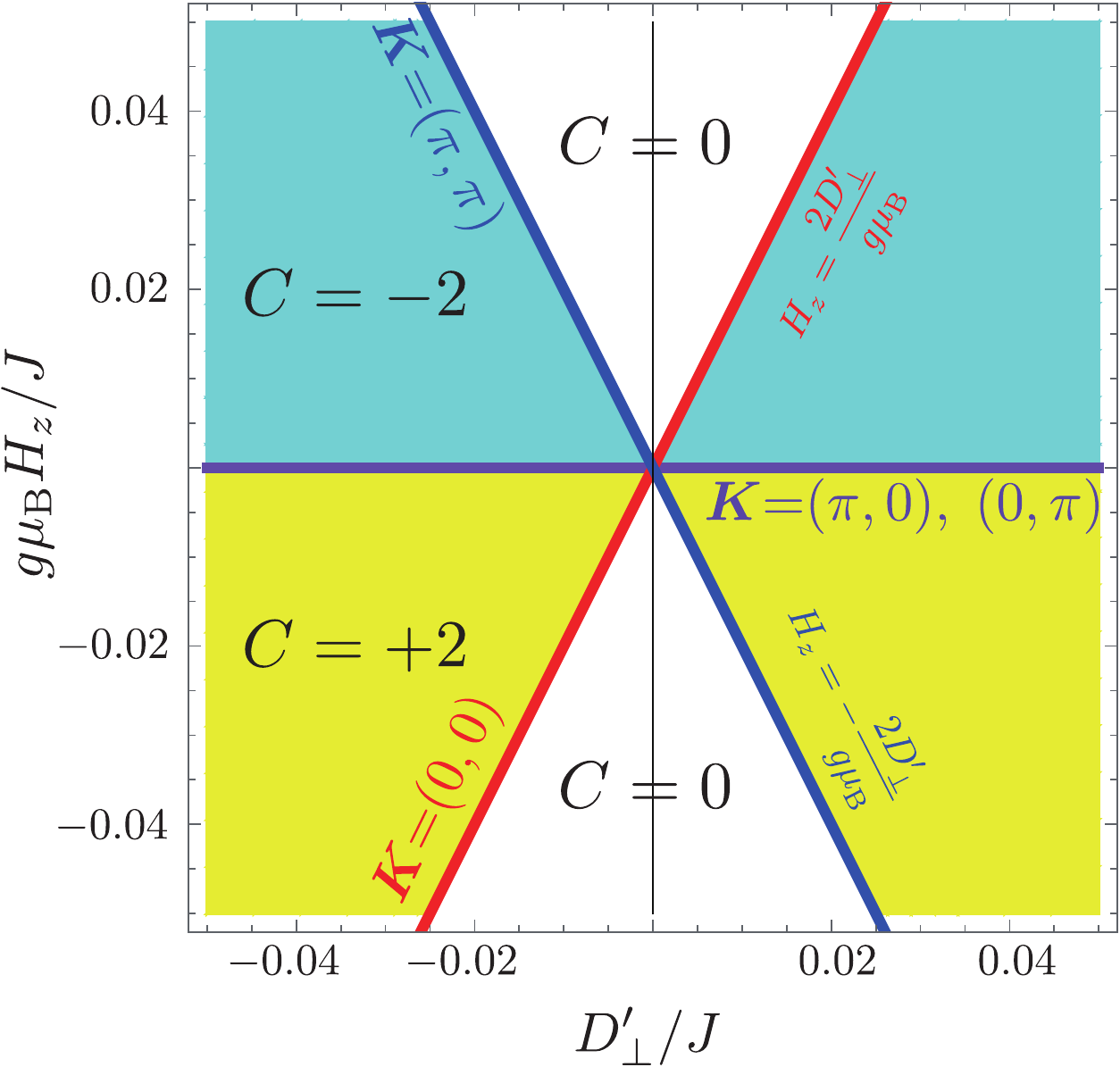}
\end{center}
\caption{Change of the triplon character in $(D^{\prime}_{\perp},H_z)$-plane.  
The Chern number $C^{(1)}$ of the top band (that of the bottom band is $-C^{(1)}$) and the positions of Dirac cones are shown.  
There are two topological ``phases'' with Chern number $C^{(1)}=+2$ (yellow) and $C^{(1)}=-2$ (cyan) as well 
as one trivial ($C^{(1)}=0$) at high fields that are separated from each other by band-touching transitions (thick lines) with Dirac cones 
(whose positions in $\bolK$-space are shown). 
\label{fig:PhaseDiag2D}}
\end{figure}

When the external field is further increased, another change in the band structure takes place at $ |H_z| = |D^{\prime}_{z}|/(2g \mu_{\text{B}})$
accompanied by a band touching with a single Dirac cone at $\bolK=(\pi,\pi)$.   
When $ |H_z| > |D^{\prime}_{z}|/(2g \mu_{\text{B}})$, $C^{(m)}= 0$ and the triplon bands lose their topological characters.  
The change in the topological character of the triplon bands is summarized in Fig.~\ref{fig:PhaseDiag2D} as a function of the field $H$.

%%%%%%%%%%%%%%%%%%%%%%%%%%%%% FIG %%%%%%%%%%%%%%%%%%%%%%%%
%%%%%%%%%%%%%%%%%%%%%%%%%%%%%%%%%%%%%%%%%%%%%%%%%%%%%%%

%%%%%%%%%%%%%%%%%%%%%%%%%%%%%%%%%%%%%%%%%%%%%%%%%%%%%%%%
\section{Thermal Hall effect from topological triplon}
\label{sec:thermal-Hall-triplon}
%%%%%%%%%%%%%%%%%%%%%%%%%%%%%%%%%%%%%%%%%%%%%%%%%%%%%%%%
In the presence of finite Berry curvature, we may generically expect the Hall response in the electrical or thermal transport \cite{Murakami-O-16}, 
and the general formula for the thermal Hall conductivity due to magnetic quasi-particles can be derived 
either by semi-classical arguments or by linear-response theory 
\cite{PhysRevLett.106.197202}.   
%%%%%%%%%%%%%%%%%%%%%%%%%%%%%%%%%%%%%%%%%%%%%%%%%%%%%%%%
\subsection{Limit of uncoupled layers}
\begin{figure*}
\includegraphics[clip,width=14cm]{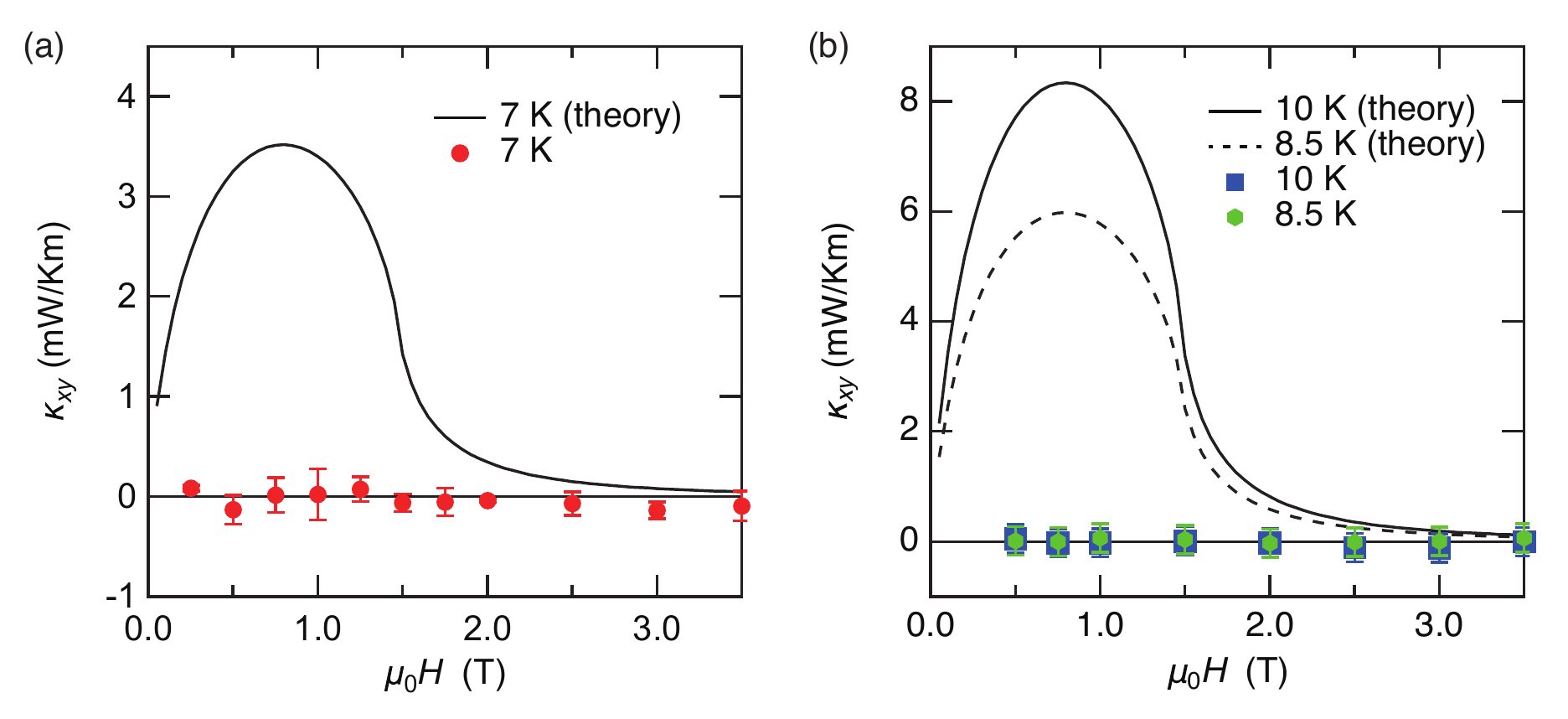}
\caption{Comparison of experimental and theoretical thermal Hall conductivities.
The magnitude of the observed $\kappa_{xy}$ (markers) is much smaller than that of the calculated $\kappa_{xy}$ by the formula given in 
Ref.~\cite{romhanyi2015hall} (solid lines).
}
\label{fig:comparison}
\end{figure*}

%%%%%%%%%%%%%%%%%%%%%%%%%%%%%%%%%%%%%%%%%%%%%%%%%%%%%%%%
When the two-dimensional Shastry-Sutherland layers (with the linear sizes $L_x \times L_y \times L_z$ and 
the number of layers $N_{\text{layer}}$) are not coupled with each other, 
the general formula given in Ref.~\cite{PhysRevLett.106.197202} leads us to the following result \cite{romhanyi2015hall}:
%%%%%%%%%%%%%%%%%%%%%%%%%%%%%%%
%\begin{widetext}
%%%%%%%%%%%%%%%%%%%%%%%%%%%%%%%
\begin{equation}
\begin{split}
& \kappa_{xy}(T,H)  \\
%& = 
%- \frac{k_{\text{B}}^{2}T}{\hbar L_x L_y L_z} \sum_{m=-1,0,+1} \sum_{\text{layer}}\sum_{\bolK=(K_x,K_y)} 
%c^{(m)}_2(\bolK ;H_z) F_{xy}^{(m)}(\bolK;H_z)   
%\\
%&= - \frac{k_{\text{B}}^{2}T}{\hbar L_x L_y L_z}  
%\frac{L_x L_y}{(2\pi)^{2}} N_{\text{layer}} \int_{\text{B.Z.}} d^{2} \bolK 
%\left\{ c^{(1)}_2(\bolK ;H_z) - c^{(-1)}_2(\bolK ;H_z) \right\}
%F_{xy}^{(1)}(\bolK ;H_z)   \\
&= - \frac{k_{\text{B}}^{2}T}{2\pi h l_z}  
\int_{\text{B.Z.}} d^{2} \bolK 
\left\{ c^{(1)}_2(\bolK ;H) - c^{(-1)}_2(\bolK ;H) \right\}
F_{xy}^{(1)}(\bolK ;H)
\end{split}
\label{eqn:triplon-kappaXY}
\end{equation}
with $l_{z}$ being the inter-layer distance and the function $c_2(\rho)$ given by:
\begin{equation}
\begin{split}
& c_2^{(m)}(\bolK;H) \equiv c_2\left( \rho_{\text{B}}(\epsilon^{(m)}(\bolK; H)) \right)    \\
& c_2(\rho) \equiv -2 \text{Li}_2(-\rho )-\log ^2(\rho )+(\rho +1) \log ^2\left(\frac{\rho +1}{\rho }\right) 
\; , \\
& \rho_{\text{B}}(\epsilon) = \frac{1}{\be^{\frac{\epsilon}{k_{\text{B}}T}} -1}   \; .
\end{split}
\end{equation} 
%%%%%%%%%%%%%%%%%%%%%%%%%%%%%%%
%\end{widetext}
%%%%%%%%%%%%%%%%%%%%%%%%%%%%%%%
All the information about the thermally excited triplons is encoded in the factor $\{ c^{(1)}_2(\bolK ;H) - c^{(-1)}_2(\bolK ;H) \}$.  
We note that, compared with the value of $\kappa_{xy}$ in the previous studies \cite{romhanyi2015hall,mcclarty2017topological}, 
the one given by Eq.~\eqref{eqn:triplon-kappaXY} is opposite in sign and smaller in magnitude by $(2\pi)^{-1}$.
While our main conclusion remains unchanged even if we used the values calculated by the formula in Ref. \cite{romhanyi2015hall} 
that further suppress the ratio $\kappa^{(\text{exp})}_{xy}/\kappa^{(\text{th})}_{xy}$ between the experimental and theoretical values 
by a factor $(2\pi)^{-1}$ (see Fig.~\ref{fig:comparison}),  
it is desirable to reexamine the previous theoretical calculations. 

A few remarks are in order here about the relation between the topology of the triplon bands and the thermal Hall transport.  
First of all, in contrast to the electrical counterpart in fermion systems, the thermal Hall response of the triplon quasi-particles 
is {\em not} quantized to the Chern number 
$\mathcal{C}^{(m)}$ (up to a constant of proportionality)  
although finite Berry curvature $F^{(m)}_{xy}(K_x,K_y)$ suggests 
finite thermal Hall transport \cite{PhysRevLett.104.066403,PhysRevLett.106.197202}.  
Moreover, the existing formula connecting the Berry curvature $F^{(m)}_{xy}(K_x,K_y)$ and the thermal Hall conductivity is derived 
by assuming the existence of the stable quasi-particles that can be treated as non-interacting.   
As will be discussed later, the geometrically-suppressed kinetic energy of triplons in $\SCBO$ 
prevents their spontaneous decay at $T$ = 0 \cite{RevModPhys.85.219,PhysRevLett.117.187203} 
that may potentially invalidate the picture based on topologically non-trivial quasi-particles. 
Therefore, this system seems to provide a unique playground to study the topological properties of bosonic quasiparticles.  

 %%%%%%%%%%%%%%%%%%%%%%%%%%%%%%%%%%%%%%%%%%%%%%%%%%%%%
 %%%%%%%%%%%%%%%%%%%%%%%%%%%%%%%%%%%%%%%%%%%%%%%%%%%%%
 \subsection{Effects of Interlayer interactions}
 \label{sec:inter-layer-appendix}
 %%%%%%%%%%%%%%%%%%%%%%%%%%%%%%%%%%%%%%%%%%%%%%%%%%%%%
 As has been discussed in Sec.~\ref{sec:inter-layer}, only the inter-layer DM interactions can crucially affect the thermal Hall conductivity. 
Symmetry analyses tell us that the possible interlayer DM interactions (on the shortest bonds connecting the upper and lower layers) 
are of the following form (see Fig.~\ref{fig:interlayer-DM}):
 \begin{equation}
\begin{split}
& \boldsymbol{d}_{1} = (- d_y, - d_y, d_z )   \; , \;\;  
\boldsymbol{d}_{2} = (- d_y,  d_y,  - d_z )  \; , \\
& \boldsymbol{d}_{3} = ( d_y, - d_y, - d_z )   \; , \;\; 
 \boldsymbol{d}_{4} = ( d_y,  d_y, d_z )    \\
 & \boldsymbol{d}^{\prime}_{1} = (- d^{\prime}_y, - d^{\prime}_y, d^{\prime}_z )   \; , \;\;  
 \boldsymbol{d}^{\prime}_{2} = (- d^{\prime}_y,  d^{\prime}_y,  - d^{\prime}_z )  \; , \\
 & \boldsymbol{d}^{\prime}_{3} = ( d^{\prime}_y, - d^{\prime}_y, - d^{\prime}_z )   \; , \;\; 
 \boldsymbol{d}^{\prime}_{4} = ( d^{\prime}_y,  d^{\prime}_y, d^{\prime}_z )  
\end{split}
\label{eqn:interlayer-DM}
\end{equation}
with $d_{y,z}$ and $d^{\prime}_{y,z}$ remaining undetermined solely by symmetry.  
Due to the buckling in the $\text{CuBO}_{3}$ ($ab$) planes, $d_{y,z}$ and $d^{\prime}_{y,z}$ are different, in general.  
The alternate stacking of the Shastry-Sutherland layers in the $c$-direction implies the period-2 structure in the stacking 
direction and the unit cell now contains four dimers [(A1, B1) for the first layer and (A2, B2) for the second].  
Correspondingly, the single-triplon hopping is now described by a $12\times 12$ matrix.  

For the interlayer DM interactions shown in Fig.~\ref{fig:interlayer-DM} and Eq.~\eqref{eqn:interlayer-DM}, 
the inter-layer part of the triplon Hamiltonian is given by:
\begin{equation}
\begin{split}
& \mathcal{H}_{\text{interlayer}} (\bolk) \\
&  =
\left( 
\begin{array}{ccccc}
\mathbf{0}_{3\times 3} & \mathbf{0}_{3\times 3} & \vline & \mathbf{0}_{3\times 3} & M_{\text{A1B2}} (\bolk) \\
\mathbf{0}_{3\times 3} & \mathbf{0}_{3\times 3} & \vline & M_{\text{A2B1}}^{\dagger} (\bolk) & \mathbf{0}_{3\times 3} \\
\hline 
\mathbf{0}_{3\times 3} & M_{\text{A2B1}} (\bolk) & \vline & \mathbf{0}_{3\times 3} & \mathbf{0}_{3\times 3} \\
M_{\text{A1B2}}^{\dagger} (\bolk) & \mathbf{0}_{3\times 3} & \vline & \mathbf{0}_{3\times 3} & \mathbf{0}_{3\times 3} \\
\end{array} 
\right)
\end{split}
\end{equation}
and 
\begin{equation}
\begin{split}
M_{\text{A1B2}} (\bolk) & \equiv 
\be^{-i \left(k_x+k_y+k_z \right)} M^{\perp}_{\text{AB}} (d_z) + \be^{-i \left(k_x+k_y \right)}M^{\perp}_{\text{AB}} (d_z^{\prime})  \\
&= 
\be^{-i \left(k_x+k_y+k_z \right)} (d_z+ d_z^{\prime} e^{i k_z} )
\begin{pmatrix}
0 & 1  & 0 \\
 - 1  & 0 & 0 \\
 0 & 0 & 0
 \end{pmatrix} 
 \\
M_{\text{A2B1}} (\bolk) & \equiv 
M^{\perp}_{\text{AB}} (d_z) + \be^{i k_z } M^{\perp}_{\text{AB}} (d_z^{\prime})  \\
& = \be^{i \left(k_x+k_y+k_z \right)} M_{\text{A1B2}} (\bolk)   
 \end{split}
 \end{equation}
 (note that the momentum $\mathbf{k}$ here is defined with respect to the original lattice and is different from $\mathbf{K}$).  
 
Although no quantitative information is available now for the values of $d_{y,z}$ and $d^{\prime}_{y,z}$ in $\SCBO$, 
we can obtain a very crude estimate by comparing the energy levels at the $\Gamma$-point with 
the ESR spectrum. Fig.~\ref{fig:ESR-12-band} plots the twelve energy levels for several values of 
the interlayer DM interaction $d_z$, where one can clearly see that finite values of $d_z$ further split the six energy levels 
found for $d_z=0$. No such splitting has been observed in ESR measurements (see Fig.~4(a) of Ref.~\cite{nojiri2003esr})  
and we may conclude 
that the interlayer DM interactions are, if exist, much smaller than those within each plane (i.e., $D_{\parallel}^{\prime}$ 
and $D^{\prime}_{\perp}$).  

For the sake of the simplicity, let us assume $d_z = d^{\prime}_z$, which allows us to work with a smaller unit cell 
(now two species of dimers A and B are identified) 
and express the interlayer hopping matrix by the following three-dimensional matrix:
\begin{equation}
\mathcal{H}_{\text{interlayer}} (\bolK) = 
\begin{pmatrix}
0 & 2i d_{z} \cos K_z & 0 \\
- 2i d_{z} \cos K_z & 0 & 0 \\
0 & 0 & 0 
\end{pmatrix}
\; .
\label{eqn:interlayer-3band}
\end{equation}
Comparing this with the 2D Hamiltonian Eq.~\eqref{eqn:3-level-model}, one sees that the net effect of the interlayer DM interactions 
is just to shift the external field in a $K_z$-dependent way:
\begin{equation}
H \to H^{\text{eff}} (K_z) = H + \frac{2d_z}{g_z \mu_{\text{B}}} \cos K_z l_z   \; , 
\end{equation}
and the summation over different layers is replaced with the $K_{z}$-integral.  
In the above expression, we have recovered the interlayer distance $l_z$ for later convenience. 
As we have the same three-level Hamiltonian \eqref{eqn:3-level-model} with $H \to H^{\text{eff}} (K_z)$, 
the Berry curvature in $(K_x,K_y)$-plane for a given $K_z$ is readily obtained just by replacing $H$ in 
Eq.~\eqref{eqn:triplon-Berry-curvature} with $H^{\text{eff}} (K_z)$.  

%%%%%%%%%%%%%%%%%%%%%%
\pagebreak
%%%%%%%%%%%%%%%%%%%%%%
\begin{widetext}
%%%%%%%%%%%%%%%%%%%%%%
Now it is straightforward to write down $\kappa_{xy}$ for 3D stack of the Shastry-Sutherland layers by 
trading $\sum_{\text{layer}}$ in Eq.~\eqref{eqn:triplon-kappaXY} for the summation over $K_z$:
\begin{equation}
\begin{split}
\kappa_{xy}^{\text{3D}} (T,H_z) & = 
- \frac{k_{\text{B}}^{2}T}{\hbar L_x L_y L_z} \sum_{m=-1,0,+1} \sum_{K_z} 
\sum_{\bolK=(K_x,K_y)} 
\tilde{c}^{(m)}_2(\bolK;K_z) F_{xy}^{(m)}(\bolK ;K_z)    \\
&= - \frac{k_{\text{B}}^{2}T}{\hbar L_x L_y L_z}  
\frac{L_x L_y L_z}{(2\pi)^{3}}  \int_{-\pi/l_z}^{\pi/l_z}\! dK_z     \int_{\text{B.Z.}} d^{2} \bolK
\left\{ \tilde{c}^{(1)}_2(\bolK;K_z) - \tilde{c}^{(-1)}_2(\bolK;K_z) \right\}
F_{xy}^{(1)}(\bolK;K_z)   \\
&= \frac{l_z}{2\pi}  \int_{-\pi/l_z}^{\pi/l_z}\! dK_z  \,  \kappa_{xy}(T,H ; K_z)  \;  ,
\end{split}
\label{eqn:triplon-kappaXY-3D}
\end{equation}
%%%%%%%%%%%%%%%%%%%%%
\end{widetext}
%%%%%%%%%%%%%%%%%%%%%
where 
\begin{equation}
\begin{split}
& \tilde{c}_2^{(m)}(\boldsymbol{K};K_z) = c_2^{(m)}(\boldsymbol{K}; H =H_{\text{eff}} (K_z))   \\
&  \kappa_{xy}(T,H ; K_z) = \kappa_{xy}(T,H=H_{\text{eff}} (K_z) )  \; .
\end{split}
\end{equation}
 When $d_z=0$, $\kappa_{xy}(T,H; K_z) $ does not depend on $K_z$ and the above 
$\kappa_{xy}^{\text{3D}} (T,H_z)$ reduces to $\kappa_{xy}(T,H_z)$ in Eq.~\eqref{eqn:triplon-kappaXY}.   

In order to further simlpilfy the expression, we trade the $K_z$-integral with the integral over the effective field:
\begin{equation*}
h_z \equiv H + \frac{2d_z}{g_z \mu_{\text{B}}} \cos K_z l_z  \quad 
\left( H - \frac{2|d_z|}{g_z \mu_{\text{B}}} \leq h_z \leq H + \frac{2|d_z|}{g_z \mu_{\text{B}}}   \right)
\; ,
\end{equation*}    
to obtain the final expression \eqref{eqn:hz-integral-for-kappa3D}:
\begin{equation}
\kappa_{xy}^{\text{3D}} (T,H) 
= \frac{1}{\pi} \int_{H  - \frac{2|d_z|}{g_z \mu_{\text{B}}}}^{H  + \frac{2|d_z|}{g_z \mu_{\text{B}}}} \! 
dh_z \frac{ \kappa_{xy}(T,h_z) }{\sqrt{\left( \frac{2d_z}{g_z \mu_{\text{B}}} \right)^{2}-(h_z -H)^{2}}}  \; .
\end{equation} 
The values obtained using this formula are shown in Fig.~\ref{fig:kappa_xy-3D}.   
%%%%%%%%%%%%%%%%%%%%%%%%%%%%%%%%%%%%%%%%%%%%%%%%%%%%%%%%

%%%%%%%%%%%%%%%%%%%%%%%%%%%%%%%%%%%%%%%%%%%%%%%%%%%%%%%%%%%%
\section{High resolution measurements of thermal Hall conductivity}
\label{sec:experimental}
\begin{figure}[H]
\begin{center}
\includegraphics[clip,width=7cm]{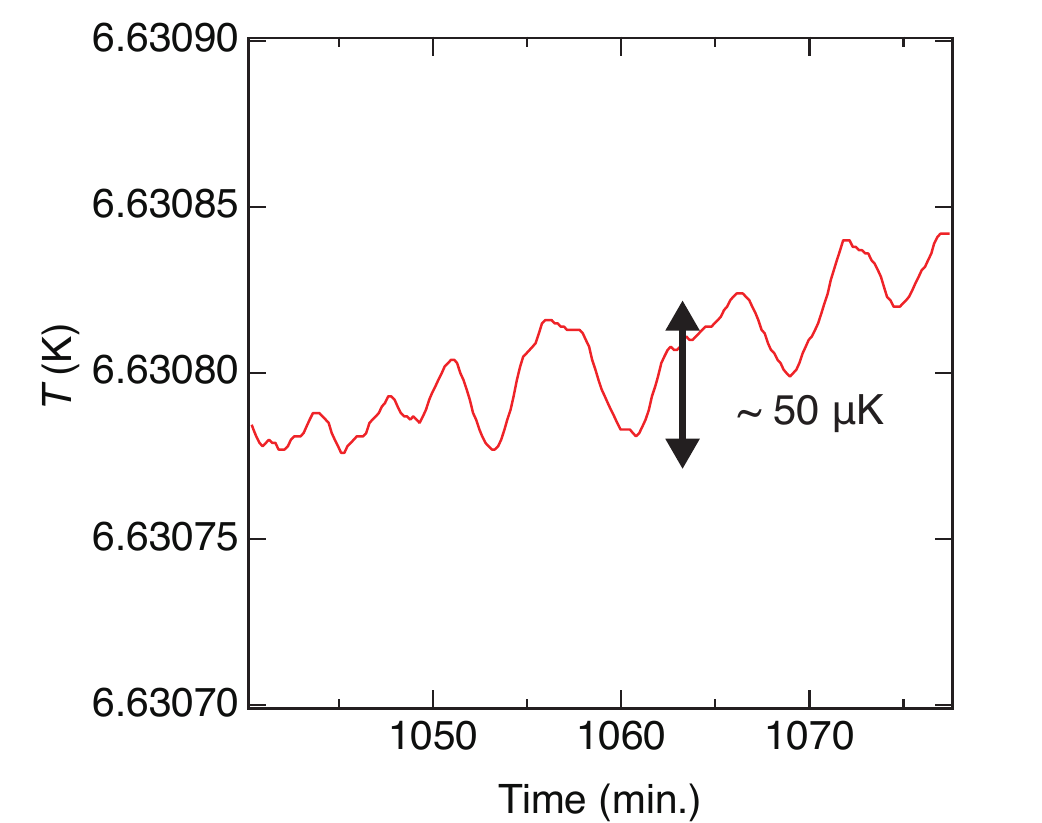}
\end{center}
\caption{Precise control of temperature during thermal Hall conductivity $\kappa_{xy}$ measurements. 
}
\label{fig:tempcontrol}
\end{figure}

\begin{figure}[H]
\begin{center}
\includegraphics[clip,width=7cm]{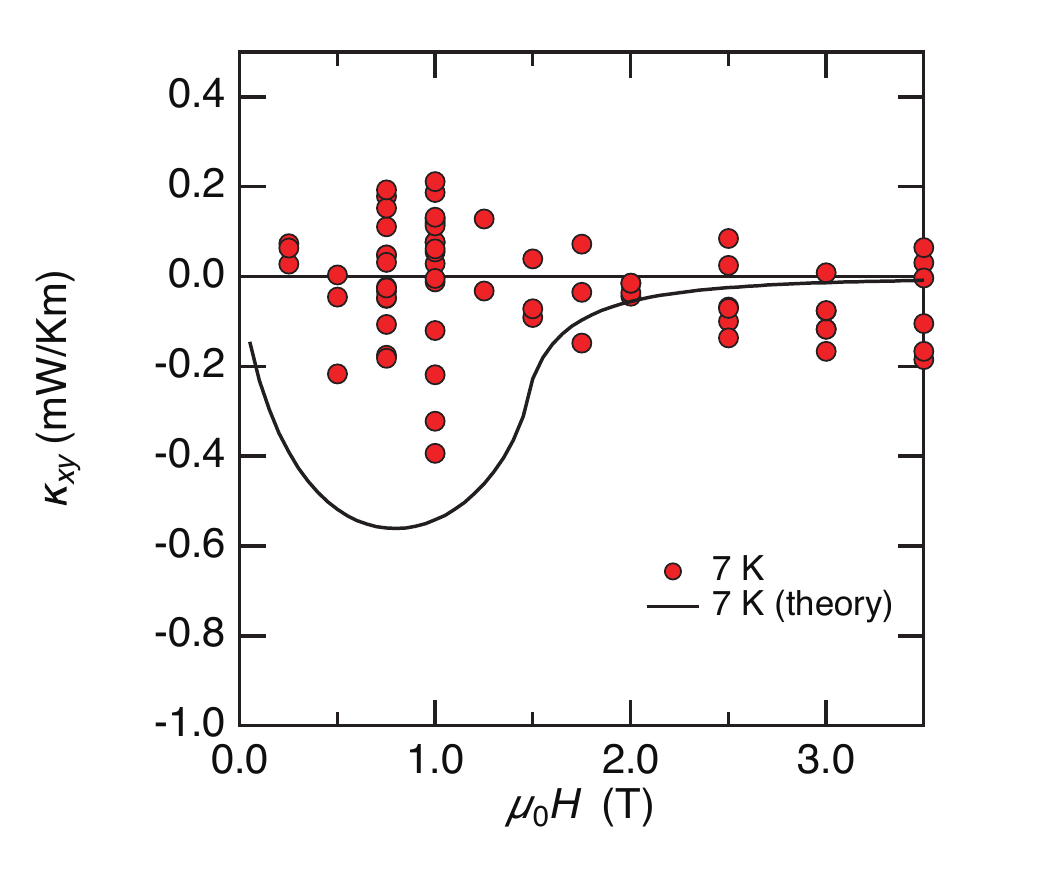}
\end{center}
\caption{Statistical error of $\kappa_{xy}$.
The error bars in Fig.\,\ref{fig:kappa_xy-comparison}(a) in the main text indicate the standard statistical errors obtained from multiple measurements (red circles).
}
\label{fig:error}
\end{figure}

%%%%%%%%%%%%%%%%%%%%%%%%%%%%%%%%%%%%%%%%%%%%%%%%%%%%%%%%%%%%
A typical temperature noise level $\delta T$ is $\sim$ 50 $\mathrm{\mu}$K at $T \sim 7 K$ ($\delta T / T \sim 10^{-5}$) (see Fig.\,\ref{fig:tempcontrol}),
which enables us to achieve high precision measurements of $\kappa_{xy}$.
In addition, we repeated the $\kappa_{xy}$ measurements more than 10 times at 0.75 and 1\,T (see Fig.\,\ref{fig:error}), where considerable $\kappa_{xy}$ of $\sim$ 0.5 mW/Km is expected.
This allows us to quantitatively compare the experimental results with the theoretical calculations (black line in Fig.\,\ref{fig:error}).
%%%%%%%%%%%%%%%%%%%%%%%%%%%%%%%%%%%%%%%%%%%%%%%%%%%%%%%%%%%%
%%%%%%%%%%%%%%%%%%%%%%%%%%%%%%%%%%%%%%%%%%%%%%%%%%%%%%%%%%%%
%%%%%%%%%%%%%%%%%%%%%%%%%%%%%%%%%%%%%%%%%%%%%%%%%%%%%%%%

% % Create the reference section using BibTeX:
\bibliographystyle{apsrev4-1}
% \bibliography{ref.bib}


\end{document}